\documentclass[aip,jmp, amsmath,amssymb,reprint, onecolumn,floatfix,nofootinbib,longbibliography]{revtex4-2}

\usepackage{graphicx}
\usepackage{silence}
\usepackage{float}
\usepackage[utf8]{inputenc}
\usepackage[T1]{fontenc}
\usepackage{etoolbox}
\usepackage{tikz-cd} 
\usepackage{hyperref}
\hypersetup{colorlinks=True,allcolors=blue}

\begin{document}

\title{On the Kaluza--Klein geometric theory in affine spaces}

\author{Oscar \surname{Castillo-Felisola}}
\email[Corresponding author: ]{o.castillo.felisola:at:proton.me}
\affiliation{Departamento de F\'isica, Universidad T\'ecnica Federico Santa Mar\'ia Casilla 110-V, Valpara\'iso, Chile.}
\affiliation{Centro Cient\'ifico Tecnol\'ogico de Valpara\'iso, Casilla 110-V, Valpara\'iso, Chile}

\author{Aureliano \surname{Skirzewski}}
\email{askirz:at:gmail.com}
\affiliation{Instituto de F\'isica, Facultad de Ciencias, Igu\'a 4225, esq. Mataojo, 11400 Montevideo, Uruguay}

\author{Jefferson \surname{Vaca-Santana}}
\email{jeffersonvaca1993:at:hotmail.com}
\affiliation{Instituto de F\'isica, Pontificia Universidad Cat\'olica de Valpara\'iso Casilla 4059, Valpara\'iso, Chile}
\affiliation{Departamento de F\'isica, Universidad T\'ecnica Federico Santa Mar\'ia Casilla 110-V, Valpara\'iso, Chile.}

\begin{abstract}
  In this work, we develop a generalization of Kaluza--Klein theory by considering a purely affine framework, without assuming a prior metric structure. We formulate the dimensional reduction using the geometry of principal fiber bundles and the Ehresmann connection, introducing adapted bases that allow an explicit decomposition of tensors, vectors, and connections. This formalism provides a natural geometric definition of the electromagnetic field as the difference between the horizontal space and the space generated by the observer’s frame. We demonstrate that the presence of a nontrivial electromagnetic field requires the non-integrability of the horizontal distribution, and we derive a complete ansatz for decomposing the affine connection into fields defined on the reduced space. Under assumptions such as vanishing torsion, autoparallel fibers, and suitable normalization conditions, we show that the reduced theory corresponds to the Einstein–Maxwell system for purely radiative electromagnetic fields. Furthermore, we propose an interpretation where the metric emerges dynamically from the affine structure through the dynamics of the electromagnetic field.
\end{abstract}

\maketitle

\section{Introduction \label{sec:Intro}}

Dimensional reduction has played a central role in theoretical physics as a tool for unifying interactions and formulating fundamental theories. A paradigmatic example is the Kaluza--Klein theory, where gravitation and electromagnetism are proposed to be unified by extending spacetime to a fifth compact dimension. This idea, introduced by Theodor Kaluza\cite{kaluza21_zum_unitaet_physik} and refined by Oskar Klein,\cite{klein26_quant_und_relat} had a lasting impact, laying the groundwork for subsequent extensions.\cite{salam82_kaluz_klein_theor,duff86_kaluz_klein_super,bailin87_kaluz,overduin97_kaluz_klein,bailin99_orbif_compac_strin}

Beyond its value as a unification mechanism, dimensional reduction also allows gauge fields and symmetries to be reinterpreted as geometric manifestations of a higher-dimensional space. This technique is widely used in modern field and gravity theories, and its correct formulation is crucial for understanding how effective interactions emerge in lower-dimensional spaces. However, most of these approaches assume the existence of a given metric on the total space, which limits their applicability to theories where the metric is a fundamental object.

In contexts where one aims to explore purely affine theories—where the connection, rather than the metric, is the primary geometric object—it becomes necessary to generalize the procedure of dimensional reduction. Although first proposals of purely affine models of gravity were due to Einstein\cite{einstein23_theor_affin_field,einstein23_zur_affin_feldt}, Eddington,\cite{eddington24_mathem_theor_relat} and developed further by Schr\"odinger,\cite{schroedinger46_general_affin_field_laws,schroedinger47_final_affin_field_laws_i,schroedinger48_final_affin_field_laws_ii,schroedinger48_final_affin_field_laws_iii,schroedinger50_space} there exist more recent proposals by Kijowski,\cite{kijowski78_new_variat_princ_gener_relat,kijowski79,ferraris81_gener_relat_is_gauge_type_theor,ferraris82_equiv_relat_theor_gravit,kijowski07_univer_affin_formul_gener_relat} Poplaski,\cite{poplawski07_nonsy_purel_affin,poplawski07_unified_purel_affin_theor_gravit_elect,poplawski09_gravit_elect,poplawski14_affin_theor_gravit} Azri,\cite{azri17_affin_inflat,azri18_cosmol_implic_affin_gravit,azri18_induc_affin_inflat,azri19_induc_gravit_from_connec_scalar_field} and Castillo-Felisola\cite{castillo-felisola15_polyn_model_purel_affin_gravit,castillo-felisola18_einst_gravit_from_polyn_affin_model,castillo-felisola18_beyond_einstein,castillo-felisola20_emerg_metric_geodes_analy_cosmol,castillo-felisola24_polyn_affin_model_gravit} and their collaborators. These are examples of frameworks seeking to construct gravitational models without assuming a metric structure \textit{a priori}, thus requiring a new geometric approach.

A natural way to address this problem is through the geometry of principal bundles and the use of the Ehresmann connection. This structure enables a local decomposition of the total space into base and fiber components, in a consistent manner and without the need for a metric. In this setting, the affine connection becomes the sole fundamental data, and a consistent dimensional reduction scheme based purely on the properties of the connection can be formulated.

This approach also has deep implications for the understanding of foliations. The Ehresmann connection allows us to interpret the reduction as a local partition of the total space into transversal leaves, and when the horizontal distribution is integrable, the resulting formalism closely resembles the Arnowitt–Deser–Misner (ADM) approach used in general relativity for the Hamiltonian analysis of spacetime.\cite{arnowitt62} This correspondence suggests a possible path toward extending canonical methods to purely affine frameworks.

The goal of this work is to develop a dimensional reduction formalism applicable to affine theories, using the structure of principal bundles and the Ehresmann connection. Within this framework, we define an explicit decomposition of vectors, tensors, and connections. A geometric definition of the electromagnetic field is introduced, and we show that, under reasonable geometric conditions, the induced field equations in the reduced space correspond to the Einstein--Maxwell equations for purely radiative fields.

The article is organized as follows. In Section \ref{sec:fibrados_conexion}, we review the fundamental concepts of principal bundle geometry and introduce the Ehresmann connection. In Sec.~\ref{sec:Descomposicion}, we present the procedure of dimensional reduction in an affine space for tensors and introduce the adapted bases. In Sec.~\ref{sec:KK-metrico}, we apply the dimensional reduction procedure to the metric case and, through comparison, analyze the emergence of an electromagnetic field and establish its geometric definition. In Sec.~\ref{sec:Descomposicion_Conexion}, we present the decomposition of the affine connection. In Sec.~\ref{sec:ReduccionCurvatura}, we compute the reduction of the Ricci tensor and show that, under certain assumptions, the resulting field equations correspond to the radiative Einstein--Maxwell system. Finally, in Section \ref{sec:Conclusiones}, we discuss the conclusions and future perspectives of this work.

\section{Fiber Bundle Geometry and Ehresmann Connection \label{sec:fibrados_conexion}}

In this section, we present the geometric foundations that support our dimensional reduction model, based on the theory of principal bundles and the Ehresmann connection. We adopt a notation inspired by the formulation of Kaluza--Klein theory, as discussed in the review article by ourselves,\cite{castillo-felisola24_polyn_affin_model_gravit} highlighting the crucial role of projection in the bundle structure.

In the classical Kaluza--Klein model,\cite{choquet-bruhat89_analy,betounes04_mathem_aspec_kaluz_klein_gravit} spacetime is described as a principal bundle, that is, a quadruple \((\widehat{\mathcal{M}}, \mathcal{M}, G, \pi)\), where:
\begin{itemize}
\item \(\widehat{\mathcal{M}}\) is a differentiable manifold called \emph{total space}.
\item \(\mathcal{M}\) is a differentiable manifold called \emph{base}.
\item \(G\) is a Lie group that acts on the right on \(\widehat{\mathcal{M}}\) via diffeomorphisms \(R_g : \widehat{\mathcal{M}} \to \widehat{\mathcal{M}}\) with \(g \in G\).
\item \(\pi : \widehat{\mathcal{M}} \to \mathcal{M}\) is a smooth map, called projector, that satisfies: 
  \begin{itemize}
  \item \(\pi(R_g(p)) = \pi(p)\) for all \(p \in \widehat{\mathcal{M}}\) and \(g \in G\) (the action preserves the fibers).
  \item The action is \emph{free}, that is, if \(R_g(p) = p\), then \(g = e\).
  \item The action is \emph{transitive on the fibers}, that is, if \(\pi(p) = \pi(q)\), then there exists \(g \in G\) such that \(q = R_g(p)\).
  \end{itemize}
\end{itemize}

Locally, for each open set \(U \subset \mathcal{M}\), one has \(\pi^{-1}(U) \cong U \times G\). The free and transitive action ensures a one-to-one correspondence between the points of each fiber and the elements of the group \(G\).

The projection \(\pi\) induces a map between tangent spaces \(T_p\widehat{\mathcal{M}}\) and \({T_{\pi(p)}\mathcal{M}}\), called the differential map (or \emph{pushforward} of \(\pi\)) and is denoted by \(\pi_*\).

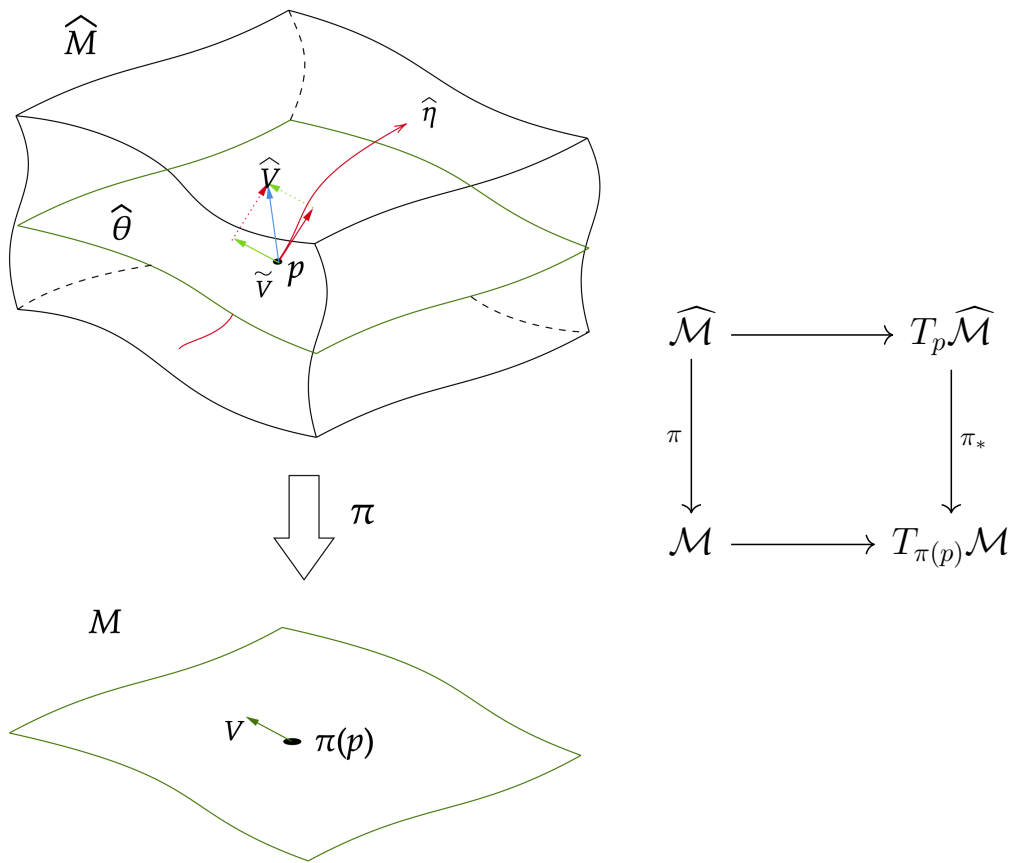
\begin{figure}[H]
  \centering
  \begin{minipage}[c]{0.58\linewidth}
    \centering
    \includegraphics[width=\linewidth]{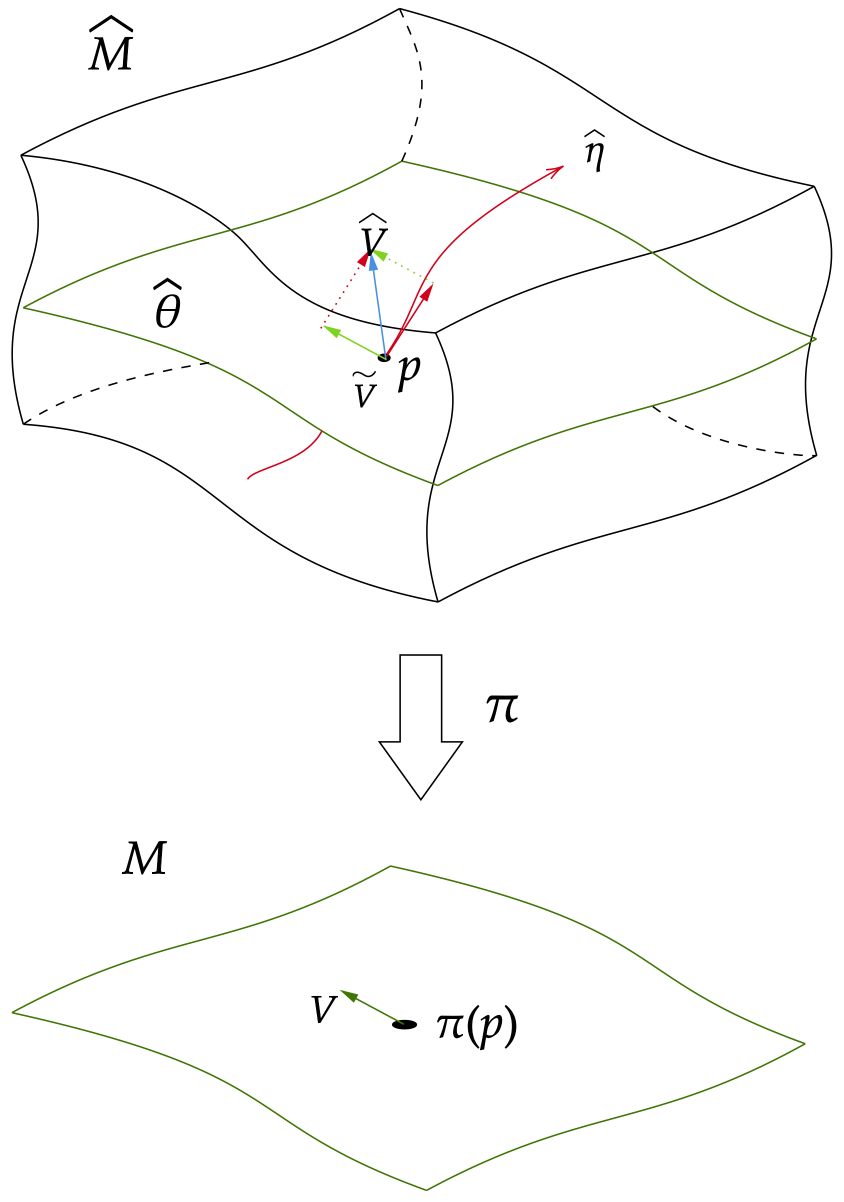}
  \end{minipage}
  \hfill
  \begin{minipage}[c]{0.4\linewidth}
    \centering
    \Large
    \begin{tikzcd}[row sep=4em, column sep=4em]
      \widehat{\mathcal{M}} \arrow[d, "\pi"'] \arrow[r] & T_p\widehat{\mathcal{M}} \arrow[d,"\pi_*"] \\
      \mathcal{M} \arrow[r] & T_{\pi(p)}\mathcal{M}
    \end{tikzcd}
  \end{minipage}
  \caption{Commuting Diagram. The figure locally illustrates the structure of a principal bundle in Kaluza--Klein theory. The total space $\widehat{\mathcal{M}}$ is projected ($\pi$) onto the base space $\mathcal{M}$ (in green). A vector $\widehat{v}$ in the tangent space $T_p\widehat{\mathcal{M}}$ is decomposed into a horizontal component $\widetilde{v}$ (green with a hat, lift of a vector in $\mathcal{M}$) and a vertical component (direction of the red vector $\widehat{\eta}$, tangent to the fiber). Under the tangent map $\pi_*$, both $\widehat{v}$ and $\widetilde{v}$ are projected to the vector $v$ in $T_{\pi(p)}\mathcal{M}$. The commuting diagram on the left summarizes this relationship between the manifolds and their tangent spaces.}
  \label{fig:bundle}
\end{figure}

The bundle structure ensures that the diagram in Fig. \ref{fig:bundle} is commutable. The commutativity is valid at every pair of points \((\widehat{p},\pi(\widehat{p}))\), continuously varying, so we omit the explicit reference to the points in the tangent spaces.

Inspired by the commutative diagram in Fig. \ref{fig:bundle}, where the projection $\pi$ acts from $\widehat{\mathcal{M}}$ to $\mathcal{M}$, and following the general convention in differential geometry of using the subscript $*$ to denote the \emph{pushforward},\cite{choquet-bruhat89_analy,nakahara03_geomet_topol_physic} as well as the usual notation in Kaluza--Klein that distinguishes between quantities in the total space  with \emph{hat} and the base space without \emph{hat}, we define the \emph{projection down} operator (\(\downarrow_{*}\)) as the pushforward of the projection operator ($\downarrow_* = \pi_*$):
\begin{equation}
  \label{eq:operador_bajada}
  \downarrow_*(\widehat{v}) = v,
\end{equation}
where $\widehat{v} \in T\widehat{\mathcal{M}}$ and $v \in T\mathcal{M}$.

Analogously, we define the \emph{co-lift} map \(\uparrow^*: T^*\mathcal{M} \rightarrow T^*\widehat{\mathcal{M}}\) such that for a covector \({L} \in T^*{\mathcal{M}}\), its lift \(\widetilde{L} \in T^*\widehat{\mathcal{M}}\) is defined  $\widetilde{L}(\widehat{v})=L(\downarrow_*(\widehat{v}))$  as follows:
\begin{equation*}
  \uparrow^*(L) = \widetilde{L}.
\end{equation*}

To perform a dimensional reduction, it is useful to locally identify a structure that allows distinguishing directions analogous to those of the base space within the total space. Due to the local nature of the representation, although in the bundle there is generally no intrinsic separation between base and fiber, it is possible to associate a subspace of the total tangent space in a way analogous to the base space. This allows decomposing the tangent space into two complementary subspaces: the vertical and the horizontal, as can be observed in Fig. \ref{fig:bundle}. This idea was formalized by Ehresmann through the introduction of a connection (also known as \emph{Ehresmann connection}), defined as a horizontal distribution \(H_p\widehat{\mathcal{M}} \subset T_p\widehat{\mathcal{M}}\) that satisfies the following properties:\cite{choquet-bruhat89_analy,nakahara03_geomet_topol_physic}
\begin{itemize}
\item \(T_p\widehat{\mathcal{M}} = H_p\widehat{\mathcal{M}} \oplus  V_p \widehat{\mathcal{M}}\), where \(V_p \widehat{\mathcal{M}} = \ker(\downarrow_*)\) is the vertical subspace and \(H_p\widehat{\mathcal{M}}\) is called \emph{horizontal subspace}.
\item The horizontal distribution is invariant under the action of \(G\), that is, for every \(p \in \widehat{\mathcal{M}}\) and \(g \in G\), one has:
  \begin{equation*}
    (R_g)_* (H_{\widehat{p}}) = H_{R_g \widehat{p} }
  \end{equation*}
\end{itemize}

This local decomposition of the tangent space is analogous to temporal or spatial foliations in Riemannian geometry. Since no additional structure is required on the bundle, this decomposition does not always extend globally, not even to a neighborhood of the point. The Frobenius theorem states that the integrability of this distribution and, therefore, the existence of the corresponding submanifold, is not guaranteed in general.\cite{schutz80_geomet_method_mathem_physic}

Alternatively, the connection can be defined by a Lie \(1\)-form:
\begin{equation*}
  \widehat{\theta} \in \Omega^1(\widehat{\mathcal{M}}, \mathfrak{g}).
\end{equation*}
That is, a \(1\)-form on \(\widehat{\mathcal{M}}\) with values in the Lie algebra \(\mathfrak{g}\). In the Kaluza--Klein context, we will show that the Ehresmann connection allows one to separate each leaf of the foliation locally, thereby identifying the base space within the total space. In the classical Kaluza--Klein model, where the fiber is isomorphic to \(U(1)\), one has \(D = \dim(\mathcal{M}) = \dim(\widehat{\mathcal{M}}) - 1\).

In our case, \(G\) is one-parameter, so \(\mathfrak{g} \cong_{\text{loc}} \mathbb{R}\) and the connection form \(\widehat{\theta}\) is a covector in \(\widehat{\mathcal{M}}\) that satisfies the following properties:
\begin{itemize}
\item \(\widehat{\theta}(X) = 0\) for every horizontal vector \(X\). In this approach, the horizontal space is the kernel of \(\widehat{\theta}\).
\item \(\widehat{\theta}(\widehat{\eta}) = 1\), where \(\widehat{\eta}\) is the fundamental vector field associated to \(\mathfrak{g}\). This is known as the first normalization condition.
\item The \(1\)-form is \(G\)-equivariant. In the case where \(G\) is one-parameter, this condition is expressed as:
  \begin{equation*}
    \pounds_{\widehat{\eta}} \widehat{\theta} = 0,
  \end{equation*}
  which indicates that \(\widehat{\eta}\) is independent of the direction of the fiber.
\end{itemize}

A vector field \(\widetilde{v} \in C^{\infty}(T\widehat{\mathcal{M}})\) is a \emph{horizontal lift} of a field \(v \in C^{\infty}(T\mathcal{M})\) if it satisfies the following,
\begin{equation}
  \label{eq:def_levantamiento_horizontal}
  \downarrow(\widetilde{v}) = v \quad \text{and} \quad \widehat{\theta}(\widetilde{v}) = 0.
\end{equation}
We denote the operator that assigns to the vector \(v \in T\mathcal{M}\) its lift \(\widetilde{v} \in C^{\infty}(T\widehat{\mathcal{M}})\) as \(\uparrow_*\), that is,
\begin{equation}
  \label{eq:operador_subida}
  \uparrow_*(v) = \widetilde{v}.
\end{equation}
In Kaluza--Klein notation, objects with a hat (such as \(\widehat{v}\)) refer to entities in higher-dimensional space, while objects without a hat (such as \(v\)) live in lower-dimensional space. We introduce the tilde notation (such as \(\widetilde{v}\)) to denote objects in a lower-dimensional space but within a higher-dimensional space.

The operator \(\uparrow_*\) establishes a point-wise isomorphism between the tangent space of the base and the horizontal subspace of the total space:
\begin{equation*}
  T_{\pi(p)}\mathcal{M} \cong H_p\widehat{\mathcal{M}} \subset T_p\widehat{\mathcal{M}}.
\end{equation*}
This isomorphism, induced by the lift and drop maps, allows one to locally relate the base space to the total space.

Furthermore, the horizontal lift allows assigning to each covector \(\widehat{L} \in T^*\widehat{\mathcal{M}}\) a covector \(L \in T^*\mathcal{M}\) such that \(L(v) = \widehat{L}(\uparrow_*(v))\). This relation defines a projection operator on the cotangent space of the total, which we call the \emph{co-projection down}:
\begin{equation}
  \downarrow^*(\widehat{L}) = L.
\end{equation}

The action of the operators \(\downarrow_*\) and \(\downarrow^*\) can be summarized (denoted) by a single operator \(\downarrow\) on tensors of any type that ``removes the hat'':
\begin{equation}
  \downarrow \widehat{T} = T,
  \label{eq:QuitarHat}
\end{equation}
where \(\widehat{T}\) and \(T\) are, respectively, sections of the tensor bundles
\begin{equation*}
  \widehat{T} \in C^{\infty}(\otimes^p T\widehat{\mathcal{M}} \otimes \otimes^q T^{*}\widehat{\mathcal{M}}), \quad \text{and} \quad T \in C^{\infty}(\otimes^p T\mathcal{M} \otimes \otimes^q T^{*}\mathcal{M}).
\end{equation*}
Analogously, the action of the operators \(\uparrow_*\) and \(\uparrow^*\) would be denoted by the operator \(\uparrow\) that ``adds a tilde'':
\begin{equation}
  \uparrow T = \widetilde{T},
\end{equation}
where \(T\) and \(\widetilde{T}\) are respectively section on the tensor bundles
\begin{equation*}
  T \in C^{\infty}(\otimes^p T\mathcal{M} \otimes \otimes^q T^{*}\mathcal{M}), \quad \text{and} \quad \widetilde{T} \in C^{\infty}(\otimes^p T\widehat{\mathcal{M}} \otimes \otimes^q T^{*}\widehat{\mathcal{M}}).
\end{equation*}

Let us define the {projection} operator \(\widehat{P}\) as the composition of the downward and upward arrows,
\begin{equation}
  \widehat{P} = \uparrow \circ \downarrow.
  \label{eq:DefProy}
\end{equation}
For simplicity, we also refer to the operation of ``adding a tilde'' as the action of \(\widehat{P}\), that is, \(\widetilde{T} = \widehat{P}(\widehat{T})\).

In this context, we distinguish three types of objects (see Table \ref{tab:notation}): hatted objects (inherent to the total space \(\widehat{\mathcal{M}}\)), unhatted objects (inherent to the reduced space \(\mathcal{M}\)), and tilde objects (horizontal lifts in \(\widehat{\mathcal{M}}\) of objects in \(\mathcal{M}\)). These distinctions are local.

\begin{table}[H]
  \centering
  \begin{tabular}{|c|c|c|}
    \hline
    \textbf{Type of Object} & \textbf{Notation} & \textbf{Description} \\ 
    \hline
    Total Space & \(\widehat{T}\) & Tensor in \(\widehat{\mathcal{M}}\) without any assumption. \\
    Reduced Space & \(T\) & Tensor in \(\mathcal{M}\) analogous to \(\widehat{T}\). \\
    Horizontal Lift & \(\widetilde{T}\) & Tensor in \(H \subset \widehat{\mathcal{M}}\) analogous to \(T\). \\ \hline
  \end{tabular}
  \caption{Summary of notation established through the article}
  \label{tab:notation}
\end{table}

For example, for a \(\binom{0}{2}\)-covariant tensor, like a metric tensor field \(g\), we have
\begin{equation}
  \begin{tikzpicture}[
    scale=1.2,
    every node/.style={scale=1},
    >={Stealth[round]},
    thick
    ]

    \node (ghat) at (0,3) {
      $\begin{gathered}
        \hat{g} : T\hat{\mathcal{M}} \times T\hat{\mathcal{M}} \to \mathbb{R} \\
        (\hat{x}, \hat{y}) \mapsto \hat{g}(\hat{x}, \hat{y})
      \end{gathered}$
    };

    \node (gtilde) at (6,3) {
      $\begin{gathered}
        \tilde{g}= \uparrow g : T\hat{\mathcal{M}} \times T\hat{\mathcal{M}} \to \mathbb{R} \\
        (\hat{x}, \hat{y}) \mapsto \tilde{g}(\hat{x}, \hat{y}) = g(x, y)
      \end{gathered}$
    };

    \node (g) at (2.8,0) {
      $\begin{gathered}
        g=\downarrow\hat{g} : T\mathcal{M} \times T\mathcal{M} \to \mathbb{R} \\
        (x, y) \mapsto g(x,y) = \hat{g}(\tilde{x}, \tilde{y})
      \end{gathered}$
    };

    \draw[->] (ghat) to[bend left] node[midway, above] {\(P\)} (gtilde);
    \draw[->] (ghat) to[bend right] node[midway, left] {$\downarrow$} (g);
    \draw[<-] (gtilde) to[bend left] node[midway, right] {$\uparrow$} (g);
  \end{tikzpicture}
\end{equation}

\section{Dimensional Decomposition and Adapted Bases \label{sec:Descomposicion}}

From the perspective of an observer restricted to \(D\) dimensions, \(\widehat{T}\) and \(\widetilde{T}\) are indistinguishable, since both are projected to the same tensor 
\begin{equation}
  T = \downarrow(\widehat{T}) = \downarrow (\widetilde{T}).
\end{equation}
However, the vectors \(\widehat{T}\) and \(\widetilde{T}\) are not necessarily equal because \(\widehat{T} - \widetilde{T} \in \ker (\downarrow)\). Hence, we can decompose the vector field \(\widehat{v} \in T\widehat{\mathcal{M}}\) as follows:
\begin{equation}
  \widehat{v} = \widetilde{v} + \widehat{A}(\widehat{v}) \widehat{\eta}.
  \label{eq:one}
\end{equation}
Thus, all the information contained in \(\widehat{v}\) is locally encoded in the pair \((\widetilde{v}, \widehat{A}(\widehat{v}))\), both defined in the reduced space.

An analogous argument applies to the covectors. For a vector \(\widetilde{v}\) in the horizontal space, one has 
\begin{equation}
  \widehat{L}(\widetilde{v}) = L(v),
\end{equation}
by the definition of the co-projection. On the other hand, by the definition of the co-lift, 
\begin{equation}
  L(v) = \widetilde{L}(\uparrow v), 
\end{equation}
so that \(\widehat{L}(\widetilde{v}) = \widetilde{L}(\widetilde{v})\), that is, when restricted to the horizontal space \(\left(\widehat{L}|_{H\widehat{\mathcal{M}}} = \widetilde{L}|_{H\widehat{\mathcal{M}}}\right)\).

In general,
\begin{equation}
  \widetilde{L}(\widehat{v}) = \uparrow \circ \downarrow \widehat{L}(\widehat{v}) = \uparrow L(v) = \widetilde{L}(\widetilde{v}),  
\end{equation}
which implies \(\widetilde{L}(\widehat{\eta}) = 0\).

In summary,
\begin{equation}
  \label{eq:CovectorL}
  L(x) = \widehat{L}(\widetilde{x}) = \widetilde{L}(\widehat{x}).
\end{equation}
Thus, a covector \(\widehat{L} \in T^*\widehat{\mathcal{M}}\) decomposes as follows,
\begin{equation}
  \widehat{L} (\widehat{v})
  = \widehat{L} (\widetilde{v}) + \widehat{L}(\widehat{\eta}) \widehat{\theta}(\widehat{v})
  = \widetilde{L} (\widehat{v}) + \widehat{L}(\widehat{\eta}) \widehat{\theta}(\widehat{v}),
\end{equation}
or equivalently,
\begin{equation}
  \widehat{L} = \widetilde{L} + \widehat{L}(\widehat{\eta}) \widehat{\theta}.
\end{equation}

This expression reveals that the covector \(\widehat{L}\) is naturally decomposed into horizontal and vertical  components, denoted by \(\widetilde{L}\) and \(\widehat{L}(\widehat{\eta})\), respectively.

Now that we know how to decompose vectors and covectors, we can follow a similar procedure to decompose higher-rank tensors. In Tab.~\ref{Tab:desc} we summarize the content of the decomposition of tensors up to rank (\(r\)) three, to a codimension one spacetime, providing also a simple analogy with a binomial of degree \(r\).

\begin{table}[H]
  \centering
  \renewcommand{\arraystretch}{1.4}
  \begin{tabular}{|c|c|c|}
    \hline
    \textbf{Rank of the} & \textbf{Components in the reduced space} & \textbf{Polynomial} \\
    \textbf{original tensor} & &  $(D+1)^n$ \\
    \hline
    1 & {1 \(1\)-tensor } + {1 \(0\)-tensor} & {\(1 D^1\)} + {\(1D^0\)} \\
    \hline
    2 & {1 \(2\)-tensor} + {2 \(1\)-tensors} + {1 \(0\)-tensor} & {\(1D^2\)} + {\(2D^1\)} + {\(1 D^0\)} \\
    \hline
    3 & {1 \(3\)-tensor} + {3 \(2\)-tensors} + {3 \(1\)-tensors} + {1 \(0\)-tensor} & {\(1D^3\)} + {\(3D^2\)} + {\(3D^1\)} + {\(1D^0\)} \\
    \hline
  \end{tabular}
  \caption{Decomposition of tensors when reducing from \(D+1\) to \(D\) dimensions. The colors indicate the correspondence between the tensor components and the terms in the polynomial.}
  \label{Tab:desc}
\end{table}

In the remaining of this section, we review some geometrical concepts, and this will serve to fix our notation.

Let \(\widehat{\mathcal{B}}\) be the smooth section of the frame bundle \(F(T\widehat{\mathcal{M}})\), defined as
\begin{equation*}
  \widehat{\mathcal{B}}_{p} = \left\{ \widehat{e}_{\mu}(p) \right\} = \left\{ \widehat{e}_{0}(p), \widehat{e}_{1}(p), \ldots, \widehat{e}_{d}(p), \widehat{e}_{D}(p) \right\}
\end{equation*}
where \(\mu = 0, 1, \cdots , D\), since \(\dim\left(T\widehat{\mathcal{M}}\right) = D+1\), and we have introduced the symbol \(d\), with value \(d = D-1\) to denote the number of \emph{spatial} dimensions of the lower dimensional manifold \(\mathcal{M}\).

Every vector field \(\widehat{v} \in T\widehat{\mathcal{M}}\) can be written as (note that hereon we shall omit the reference to the point on the manifold),
\begin{equation*}
  \widehat{v} = \widehat{v}^{\nu} \widehat{e}_{\nu},
\end{equation*}
where \(\widehat{v}^{\nu}\) denotes the components of the vector field in the basis defined by the section \(\widehat{\mathcal{B}}\). The corresponding dual basis is determined by the smooth section \(\widehat{\mathcal{B}}^{*}\) of the frame bundle \(F(T^{*}\widehat{\mathcal{M}})\), defined as
\begin{equation*}
  \widehat{\mathcal{B}}^{*} = \left\{ \widehat{e}^{\mu} \right\} = \left\{ \widehat{e}^{0}, \widehat{e}^{1}, \ldots, \widehat{e}^d, \widehat{e}^D \right\}.
\end{equation*}

The question now is: How does the section \(\widehat{\mathcal{B}} \in C^{\infty}(F(T\widehat{\mathcal{M}}))\) (equivalently \(\widehat{\mathcal{B}}^{*} \in C^{\infty}(F(T^{*}\widehat{\mathcal{M}}))\)) induce a well-defined smooth section \(\mathcal{B}\) on the frame bundle \(F(T\mathcal{M})\)? The answer depends on the behavior of the vectors under the projection \(\pi_*\), and we can distinguish two cases:
\begin{enumerate}
\item All vectors project to a non-zero element:
  
  In this case, each \(\widehat{e}_\mu\) has a non-zero projection. In particular, the last vector can be written as
  \begin{equation*}
    \widehat{e}_{D} = \widehat{B} + \phi \eta,
  \end{equation*}
  where \(\widehat{B} \in \ker(\widehat{\theta})\) and \(\phi \ne 0\). The induced basis in \(T\mathcal{M}\) is formed by the vectors
  \begin{equation*}
    e_i = \downarrow(\widehat{e}_i), \quad \text{for } i = 0, \ldots, d,
  \end{equation*}
  and it is \emph{unique}: the projected vectors preserve the order and form a basis of the reduced space.
  
\item Some vector vanishes under the projection:
  
  Suppose there exists \(\nu\) such that \(\pi_*(\widehat{e}_\nu) = 0\). Without loss of generality, we can reorder the basis by placing that vector at the end, so that
  \begin{equation*}
    \pi_*(\widehat{e}_{D}) = 0 \quad \Rightarrow \quad \widehat{e}_{\widehat{d}} = \phi \eta, \quad \phi \ne 0.
  \end{equation*}
  Then, the vectors \(\widehat{e}_i\) with \(i = 0, \ldots, d\) have non-zero projections, and the induced basis in \(T\mathcal{M}\) is
  \begin{equation*}
    e_i = \downarrow(\widehat{e}_i).
  \end{equation*}
  Again, the order is clear and \emph{unique}, since the vector that projects to zero has been placed at the end.
\end{enumerate}

In summary, in both cases a unique basis in \(T\mathcal{M}\) can be induced simply by projecting the set of the first \(D\) vector fields from the frame \(\widehat{\mathcal{B}}\). Thus, we define 
\begin{equation*}
  \mathcal{B} = \left\{ e_{0},e_{1},...,e_{d}\right\},
\end{equation*}
such that every smooth section \(v \in C^{\infty}(T\mathcal{M})\) might be expressed as $v=v^{i}e_{i}$, where the Latin indices take values from \(0\) to \(d\). And its respective dual basis is determined by a section of the frame bundle \(F(T\mathcal{M})\), given by
\begin{equation*}
  \mathcal{B}^{\ast}=\left\{ e^{0},e^{1},...,e^{d}\right\}.
\end{equation*}

Since there exists a bijection between $T\mathcal{M}$ and the horizontal lift \(H\widehat{\mathcal{M}}\), the basis \(\mathcal{B}\) induces a basis in the horizontal lift given by \(\widehat{E}_i = \widetilde{e}_i = \uparrow e_i\). However, given that \(H\widehat{\mathcal{M}} \subset T\widehat{\mathcal{M}}\), we might consider another section \(\widetilde{\mathcal{B}}\) on the frame bundle \(F(T\widehat{\mathcal{M}})\) using the natural basis on the horizontal lift and complementing it with the vector along the vertical bundle, i.e. \(\widehat{E}_D = \widehat{\eta}\), inducing an adapted basis to the projection 
\begin{equation*}
  \widetilde{\mathcal{B}} = \left\{ \widehat{E}_{0}, \widehat{E}_{1}, ..., \widehat{E}_{D}\right\}=\left\{ \widetilde{e}_{0},\widetilde{e}_{1},...,\widetilde{e}_{d},\widehat{\eta}\right\},
\end{equation*}
and its respective dual basis
\begin{equation*}
  \widetilde{\mathcal{B}}^{\ast} = \left\{ \widehat{E}^{0}, \widehat{E}^{1}, ..., \widehat{E}^{D} \right\}.
\end{equation*}

Note that a vector \(\widehat{v} \in T\widehat{\mathcal{M}}\) in the basis \(\widetilde{\mathcal{B}}\) is written as
\begin{equation*}
  \widehat{v} = \widetilde{v}^{\mu} \widehat{E}_{\mu} = \widetilde{v}^{i} \widetilde{e}_{i} + \widetilde{v}^{D} \widehat{\eta} = \widetilde{v}^{i}\widetilde{e}_{i} + \theta(\widehat{v}) \widehat{\eta},
\end{equation*}
where we have defined \(\widetilde{v}^{D} = \theta(\widehat{v})\). With this notation, on the adapted basis \(\widetilde{\mathcal{B}}\), a vector \(\widehat{v}\) belongs to the horizontal space if and only if \(\widetilde{v}^{D} = 0\).

The relation between the components of the frame fields \(\widehat{\mathcal{B}}\) and \(\widetilde{\mathcal{B}}\), and \(\widehat{\mathcal{B}}^{*}\) and \(\widetilde{\mathcal{B}}^{*}\) is given by
\begin{equation}
  \label{eq:components-relations}
  \begin{aligned}
    \widehat{e}_i & = \widehat{E}_i+\widehat{\theta}_i \widehat{E}_D, & \widehat{e}_D & = \widetilde{B} ^i \widehat{E}_i+\phi \widehat{E}_D, \\
    \widehat{E}^k & =  \widehat{e}^k + \widetilde{B}^k  \widehat{e}^D, & \widehat{E}^D & = \widehat{\theta}
  \end{aligned}
\end{equation}

In the next section, we show how a rank two tensor, such as the metric, is decomposed. Then, in the following section, an analogous procedure will be used to decompose the connection.

\section{Affine perspective of the Kaluza--Klein dimensional reduction \label{sec:KK-metrico}}

In this section, we reproduce the Kaluza--Klein metric decomposition and propose a geometric definition of the electromagnetic field.

Our model starts considering a bundle \((\widehat{\mathcal{M}}, \mathcal{M}, \pi, G)\), where the total manifold would be the higher-dimensional spacetime, the base manifold the lower-dimensional spacetime, and the fiber corresponds to the extra dimensions. We further assume that the total manifold is endowed with a metric tensor field \(\widehat{g} \in C^{\infty}( T\widehat{\mathcal{M}} \otimes T\widehat{\mathcal{M}})\).

The metric tensor field \(\widehat{g}\) can be decomposed using projections induced by the map \(\pi\), resulting in a metric tensor field \(g = \downarrow \widehat{g}\), two covectors \(\widehat{\alpha}_1(\widehat{x}) = \widehat{g}(\widehat{x},\widehat\eta)\) and \(\widehat{\alpha}_2(\widehat{x}) = \widehat{g}(\eta,\widehat{x})\), and a scalar \(\widehat{\phi} = \sqrt{ \widehat{g}(\widehat\eta,\widehat\eta)}\), in the base manifold. Due to the symmetry of $\widehat{g}$, we have \(\widehat{\alpha}_1 = \widehat{\alpha}_2 = \widehat{\alpha}\).

If we further assume that the vertical space $V\widehat{\mathcal{M}}$ is orthogonal to the horizontal space $H\widehat{\mathcal{M}}$, then $H\widehat{\mathcal{M}} = \ker(\widehat{\alpha})$, which implies that $\widehat{\alpha} = \widehat{\phi}^2 \widehat{\theta}$ and $\widehat{g}(\widetilde{x},\widehat\eta)=0$, allowing the decomposition of the metric as follows
\begin{align*}
  \widehat{g}(\widehat{x},\widehat{y})
  & = \widehat{g}(\widetilde{x}+\widehat{\theta}(\widehat{x})\widehat\eta,\widetilde{y}+\widehat{\theta}(\widehat{y})\widehat\eta)
  \\
  & = \widehat{g}(\widetilde{x},\widetilde{y}) + \widehat{\theta}(\widehat{x}) \widehat{\theta}(\widehat{y}) \widehat{\phi} ^2,
\end{align*}
or in tensor notation,
\begin{equation}
  \widehat{g} = \widetilde{g} + \widehat{\theta} \otimes \widehat{ \theta} \, \widehat{\phi}^2.
\end{equation}

Expressing the above in the basis $\mathcal{B} = \left\{ \widehat{e}_{0}, \widehat{e}_{1}, \cdots, \widehat{e}_{D} =  \widehat{\eta}\right\}$, we recover the decomposition
\begin{equation}
  \widehat{g}  =
  \left(
  \begin{array}{c|c}
    \widehat{g}_{ij} & \widehat{g}_{iD} \\
    \hline
    \widehat{g}_{Dj} & \widehat{g}_{DD}
  \end{array}
  \right)
  = 
  \left(
  \begin{array}{c|c}
    \widetilde{g}_{ij} + \widehat{\theta}_{i} \widehat{\theta}_{j} \widehat{\phi}^2 & \widehat{\theta}_{i} \widehat{\phi}^2 \\
    \hline
    \widehat{\theta}_{j} \widehat{\phi} ^2 & \widehat{\phi}^2
  \end{array}
  \right),
\end{equation}
which is the ansatz usually used in Kaluza--Klein theory that has been obtained using the affine projection formalism. Thus, we can note that although we initially assume that \(\widehat{\theta}\) is independent of the metric, we find a relationship between \(\widehat{\theta}\) and the metric.


Observe that we require the basis used to satisfy the particularity
\begin{equation*}
  \widehat{e}_D = \widehat{\eta}.
\end{equation*}
Since in the adapted bases, $\widehat{B}=0$ and $\phi=1$, from here on we shall assume these constraints,
\begin{equation}
  \widehat{B}=0 \quad  \phi=1 .
  \label{D-1Condicion}
\end{equation}

Usually in the dimensional reduction a la Kaluza--Klein, the vector potential of the electromagnetic field is defined as
\begin{equation}
  A = \widehat{\theta}_i e^i.
\end{equation}
If we do the horizontal lift in this way, it is $\widetilde{A} = \widehat{\theta}_i \widetilde{e}^i $. From Eq.~\eqref{eq:components-relations}, we have that
\begin{equation}
  \widehat{E}^D =  \widehat\theta = \widehat{\theta}_i \widehat{e}^i + \widehat{\theta}_D \widehat{e}^D = \widetilde{A}+  \widehat{e}^D.
\end{equation}
Thus, we can define the lift of the electromagnetic field as
\begin{equation}
  \widetilde{A} = \widehat{E}^D -  \widehat{e}^D
  \label{A_Def}
\end{equation}

It is important to note that $\widetilde{A}_i=\widehat{\theta}_i= A_i$, but that $\widehat{\theta}_D=1$ while $\widetilde{A}_D=0$.

Under a change of coordinates, $\widehat{x}'^\mu(\widehat{x}^i , \widehat{x}^D)$, the bases transform according to the rule
\begin{equation}
  \widehat{e}_\mu= \widehat\partial_\mu=\frac{\partial  \widehat{x}'^\nu}{\partial \widehat{x}^ \mu} \partial'_\nu= \frac{\partial  \widehat{x}'^\nu}{\partial \widehat{x}^ \mu} \widehat{e}'_\nu.
\end{equation}
Note that for the index $\mu =D$, we would like to preserve equation \eqref{D-1Condicion}, and therefore $\widehat{e}_D=\widehat{e}'_D$. Hence,
\begin{equation}
  \begin{aligned}
    \frac{\partial \widehat{x}'^ D}{\partial \widehat{x}^ D} & = 1 & \Rightarrow & \widehat{x}'^ D= \widehat{x}^D +\lambda(x^i) \\
    \frac{\partial  \widehat{x}'^j}{\partial x^ D} & = 0 & \Rightarrow & \widehat{x}'^j=\widehat{x}'^j (x^i).
  \end{aligned}
\end{equation}

For the dual basis, it follows that
\begin{equation}
  \widehat{e}'^\mu=d\widehat{x}'^\mu=\frac{\partial \widehat{x}'^ \mu }{\partial \widehat{x}^\nu}dx^\nu = \frac{\partial \widehat{x}'^ \mu }{\partial \widehat{x}^\nu}  \widehat{e}^\nu.
\end{equation}
Considering the case $\widehat{x}^{\prime j} = \widehat{x}^j$, we see that
\begin{align}
  \widehat{e}'^j & = \widehat{e}^j ,
  \\
  \widehat{e}'^D & = \widehat{e}^D + \partial_i \lambda \, \widehat{e}^i.
\end{align}

The invariance of Eq.~\eqref{A_Def} requires the following:
\begin{equation}
  \label{eq:gauge-transformation-A}
  \begin{aligned}
    \widehat{A}' & = \widehat{\theta} - \widehat{e}'^{D}\\
    \widehat{A}' & = \widehat{\theta} - \widehat{e}^D- \partial_i \lambda \widehat{e}^i\\
    \widehat{A}'+ \partial_i \lambda \widehat{e}^i & = \widehat{\theta} - \widehat{e}^D
  \end{aligned}
\end{equation}
We have to define $\widehat {A}= \widehat{A}'+ \partial_i \lambda \widehat{e}^i $, which implies that $A_i=A'_i+\partial_i \lambda$, as expected, complies with the gauge transformation of the electromagnetic potential.\cite{jackson99_class_elect}

It is worth highlighting that the manipulations that yield the electromagnetic potential do not require a metric. According to Eq.~\eqref{A_Def}, it is defined as the difference of the choices of the \((D+1)\)-th base vector in the frame \(\widehat{\mathcal{B}}\), and the one used to complement the adapted basis to the horizontal space (as subspace of the tangent of \(\widehat{\mathcal{M}}\)). This geometric formulation provides a natural interpretation of gauge transformations and highlights that the definition of the electromagnetic field does not depend on a metric a priori, but rather on the structure of the underlying bundle.

An interesting consequence of equation \eqref{eq:components-relations}, where $\widehat{e}_i = \widehat{E}_i + \widetilde{A}_i \widehat{\eta}$, can be obtained by calculating the Lie bracket between these fields, that is, \([\widehat{e}_i, \widehat{e}_j]\).

Consider a local chart on \(\widehat{\mathcal{M}}\) with coordinates $(\widehat{x}^i,\widehat{x}^D)$, where $\widehat{x}^D$ parameterizes the fiber. The functions $\widehat{x}^i$ induce the coordinate system in the base
\begin{equation*}
  x^i = \pi^i(\widehat{x}^j,\widehat{x}^D).
\end{equation*}
In $\widehat{\mathcal{M}}$ the fields $\widehat{\partial}_i = \partial/\partial\widehat{x}^i$ are generally non horizontal. By the definition of the Ehresmann connection, the horizontal lift $\widehat{E}_i$ of $\partial_i=\partial/\partial x^i$ satisfies
\begin{equation*}
  \widehat{\partial}_i = \widehat{E}_i + A_i\,\widehat{\partial}_D,
\end{equation*}
with $A_i=A_i(x^j)$. It then follows that $\widehat{\partial}_i\neq\widehat{E}_i$ unless it acts on independent functions of $\widehat{x}^D$, in which case $\widehat{\partial}_i f(x^j)=\partial_i f(x^j)$. This hypothesis of a \emph{transitive} connection legitimizes the use of $\partial_i$ (without a hat) for these fields.

It is important to distinguish between partial derivatives as operators and vectors coordinated in the tangent bundle. In fact, the projection
\begin{equation*}
  \downarrow\,\widehat{e}_i = \downarrow\,\widehat{\partial}_i = \downarrow\,\widehat{E}_i = \partial_i
\end{equation*}
must be understood as an equality between \emph{vector fields} in $\mathcal{M}$, not a mere notational convention.

The natural basis $\{\widehat{e}_i=\partial/\partial\widehat{x}^i\}$ is integrable, which implies
\begin{equation*}
  [\widehat{e}_i,\widehat{e}_j] = 0.
\end{equation*}
Then, using the decomposition $\widehat{e}_i=\widehat{E}_i+\widetilde{A}_i\,\widehat\eta$, it follows that
\begin{equation*}
  [\widehat{e}_i,\widehat{e}_j]
  = [\widehat{E}_i,\widehat{E}_j]
  + \widetilde{A}_j[\widehat{E}_i,\widehat\eta]
  - \widetilde{A}_i[\widehat{E}_j,\widehat\eta]
  + (\widehat{E}_i\widetilde{A}_j - \widehat{E}_j\widetilde{A}_i)\,\widehat\eta.
\end{equation*}
Since $\widehat{E}_i\widetilde{A}_j=(\partial_i - A_i\partial_D)A_j =\partial_iA_j $ and $[\widehat{E}_i,\widehat\eta]= -\,\partial_DA_i\,\widehat\eta=0$, we have 
\begin{equation*}
  [\widehat{E}_i,\widehat{E}_j] = -\,F_{ij}\,\widehat\eta,
  \qquad
  \widehat{\theta}\bigl([\widehat{E}_i,\widehat{E}_j]\bigr) = - F_{ij}.
\end{equation*}

Now, consider the integrability condition of the horizontal distribution $H\widehat{\mathcal{M}}$. Frobenius' theorem states that the horizontal distribution $H\widehat{\mathcal{M}}$ is integrable\footnote{Frobenius' integrability means that, there exists a subvariety $\widetilde{\mathcal{M}} \subset \widehat{\mathcal{M}}$ such that $T_p\widetilde{\mathcal{M}} = H_p\widehat{\mathcal{M}}$ for all $p \in \widetilde{\mathcal{M}}$, analogous to the base space as locally illustrated in Figure \ref{fig:bundle}.} if and only if the Lie bracket of any two vector fields that belong to $H\widehat{\mathcal{M}}$ also belongs to $H\widehat{\mathcal{M}}$:
\begin{equation*}
  [\widehat{X}, \widehat{Y}] \in H\widehat{\mathcal{M}}, \quad \forall \widehat{X}, \widehat{Y} \in H\widehat{\mathcal{M}}.
\end{equation*}
Since the fields $\widehat{E}_i$ form the basis of the horizontal space, the integrability condition translates into the fact that the commutator of two of these fields, $[\widehat{E}_i, \widehat{E}_j]$, must be horizontal. A vector is horizontal if and only if it vanishes under the action of the connection form $\widehat{\theta}$. Therefore, the integrability condition of the horizontal distribution $H\widehat{\mathcal{M}}$ is expressed as:
\begin{equation}
  \widehat{\theta}([\widehat{E}_i, \widehat{E}_j]) = 0, \text{ or equivalently } F_{ij} = 0.
\end{equation}

Therefore, if the horizontal distribution is integrable, the electromagnetic field must be trivial. This is a generalization of the result obtained by Li-Xin Li.\cite{li23_note_kaluz_klein_theor} Thus, the only way to have a non-null electromagnetic field that is consistent with a geometric theory based on bundles is through the non-integrability of the horizontal distribution.

\section{Connection Decomposition \label{sec:Descomposicion_Conexion}}

In this section, we consider an affine connection $\widehat{\nabla}$ defined in the total space $\widehat{M}$, compatible with the symmetry of the principal bundle induced by the action of a one-parameter Lie group $G$. In particular, we require the connection to be $G$-invariant,\cite{bleecker81_gauge_theor_variat_princ,choquet-bruhat89_analy} which means it satisfies
\begin{equation*}
  (R_g)_* (\widehat{\nabla}_{\widehat{X}} \widehat{Y}) = \widehat{\nabla}_{(R_g)_* \widehat{X}} ((R_g)_* \widehat{Y}),
\end{equation*}
for all $g \in G$ and for all pairs of vector fields $\widehat{X},\widehat{Y} \in C^\infty(T^*\widehat{M})$.

This condition reflects the commutation between the covariant derivation of a field and the push-forward by the action $R_g$.

Working in a local chart, with $\widehat{\eta} = \widehat{\partial}_D$, the above condition translates into a restriction on the coefficients of the affine connection,\cite{choquet-bruhat89_analy,nakahara03_geomet_topol_physic}
\begin{equation}
  \partial_D \widehat{\Gamma}^\rho_{\mu\nu} = 0,
  \label{eq:GInvarianzaConexion}
\end{equation}
i.e., the coefficients of the affine connection must be constant along the fibers generated by $\widehat{\eta}$. From a geometric perspective, this reflects the invariance of the connection under the flow generated by $G$, therefore, its compatibility with the structure of the bundle.

The decomposition of the affine connection is based on projecting the covariant derivatives from the total space $\widehat{\mathcal{M}}$ onto the base space $\mathcal{M}$, using the Ehresmann connection and the structure of the principal bundle. Our strategy is to systematically analyze the action of $\widehat{\nabla}$ on vector fields belonging to the horizontal and vertical distributions. Since any vector field $\widehat{X} \in T\widehat{\mathcal{M}}$ can be decomposed into its horizontal part $\widetilde{X}$ and its vertical part (proportional to $\widehat{\eta}$), we only need to consider the covariant derivatives of the basis fields $\{\widetilde{e}_i, \widehat{\eta}\}$. The result of each derivative, which is a vector in $T\widehat{\mathcal{M}}$, can be projected directly onto the horizontal and vertical subspaces. This procedure naturally defines a complete set of tensor fields on the base manifold $\mathcal{M}$, which fully characterize the original connection. This strategy is consistent with the general decomposition of the tensors presented in Table \ref{Tab:desc}.

Let $X, Y$ be vector fields on the base manifold $\mathcal{M}$, and $\widetilde{X}, \widetilde{Y}$ be their horizontal lifts. We analyze the four fundamental operations:
\paragraph{Transport of horizontal vectors along horizontal directions, \(\widehat{\nabla}_{\widetilde{X}}\widetilde{Y}\).---} This is the most direct analogue to transport on the base manifold. Its horizontal projection defines the \emph{induced connection} $\nabla$ on $\mathcal{M}$, while its vertical projection defines a \((0,2)\)-tensor field $h$,
\begin{align}
  \nabla_X Y &= \downarrow (\widehat{\nabla}_{\widetilde{X}} \widetilde{Y}), \label{eq:nabla_def} \\
  h(X, Y) &= \widehat{\theta}(\widehat{\nabla}_{\widetilde{X}} \widetilde{Y}). \label{eq:h_def}
\end{align}
The object $\nabla$ inherits the properties of an affine connection. The tensor $h$ measures the extent to which the horizontal space is not transported in parallel along itself, acting as an extrinsic curvature or the second fundamental form.
\paragraph{Transport of the vertical vector along horizontal directions, \(\widehat{\nabla}_{\widetilde{X}}\widehat{\eta}\).---}
This describes how the direction of the fiber changes as one moves across the base. Its horizontal part defines a \((1,1)\)-tensor $\beta$, and its vertical part defines a \(1\)-form $\tau$:
\begin{align}
  \beta(X) &= \downarrow(\widehat{\nabla}_{\widetilde{X}} \widehat{\eta}), \label{eq:beta_def} \\
  \tau(X) &= \widehat{\theta}(\widehat{\nabla}_{\widetilde{X}} \widehat{\eta}). \label{eq:tau_def}
\end{align}
The tensor $\beta$ measures the \emph{tilting} of the fiber relative to the horizontal distribution, while $\tau$ measures its \emph{stretching} (or shrinking).
\paragraph{Transport of horizontal vectors along the vertical direction, \(\widehat{\nabla}_{\widehat{\eta}}\widetilde{Y}\).---}
This captures how horizontal vectors vary along the fibers. Since this operation is not tensorial in $Y$, we must subtract the Lie derivative term to define a proper tensor $\gamma$. The decomposition is given by
\begin{align}
  \gamma(Y) &= \downarrow(\widehat{\nabla}_{\widehat{\eta}} \widetilde{Y} - \pounds_{\widehat{\eta}} \widetilde{Y}), \label{eq:gamma_def} \\
  w(Y) &= \widehat{\theta}(\widehat{\nabla}_{\widehat{\eta}} \widetilde{Y}). \label{eq:w_def}
\end{align}
The tensor $\gamma$ captures the covariant change of horizontal vectors along the fiber, while the \(1\)-form $w$ is its vertical component.

Although we might be tempted to define $\gamma(X) = \pi_*(\widehat{\nabla}_{\widehat{\eta}} \widetilde{X})$, this expression does not define a tensor, since $\widehat{\nabla}$ is not a tensorial operator. Compared with the definition of $\beta$, it becomes evident that it is necessary to subtract the non-tensorial part, which justifies the appearance of the Lie term.
\paragraph{Transport of the vertical vector along itself, \(\widehat{\nabla}_{\widehat{\eta}}\widehat{\eta}\).---}
This represents the self-acceleration of the fiber. Its horizontal component defines a vector field $\sigma$, while its vertical component defines a scalar field $\psi$:
\begin{align}
  \sigma &= \downarrow(\widehat{\nabla}_{\widehat{\eta}} \widehat{\eta}), \label{eq:sigma_def} \\
  \psi &= \widehat{\theta}(\widehat{\nabla}_{\widehat{\eta}} \widehat{\eta}). \label{eq:psi_def}
\end{align}
The vector $\sigma$ describes the projection of this self-acceleration onto the base , and $\psi$ measures its component purely along the fiber.

Of these eight fields, the induced connection $\nabla$ requires special treatment. As it is not a local tensor field, no direct horizontal lift is applied to it. Instead, its components $\Gamma_i{}^k{}_j$ are obtained from the total connection $\widehat{\nabla}$. There are two equivalent definitions for this procedure.

On the one hand, the components of the affine connection in \(\mathcal{M}\) can be obtained directly from the projection of the covariant derivative in \(\widehat{\mathcal{M}}\), through the relation 
\begin{equation}
  \Gamma_i{}^k{}_j = e^k ( \downarrow (\widehat{\nabla}_{\widetilde{e}_i} \widetilde{e}_j) ).
\end{equation}

On the other hand, a useful alternative consists in reinterpreting the connection through the \(1\)-form $\widehat{\Gamma}^\kappa{}_\nu = \widehat{\Gamma}_{\mu}{}^\kappa{}_\nu \widehat{e}^\mu$. The procedure consists of projecting each of these forms onto the horizontal subspace to obtain $\widetilde{\Gamma}^\kappa{}_\nu = \widehat{P} \widehat{\Gamma}^\kappa{}_\nu$, and subsequently projecting their indices. This leads to the following expression for the connection form of the base space:
\begin{equation}
  \Gamma^k{}_j = \widehat{P}^k{}_\kappa \widehat{P}^\nu{}_j \, \downarrow^* \widetilde{\Gamma}^\kappa{}_\nu. \label{eq:AltProjectedConnection}
\end{equation}
In Appendix~\ref{Calculations_Connection_Decomposition} we demonstrate the equivalence of both approaches.

In contrast, the other seven fields that complete the decomposition $(\nabla, h, \beta, \gamma, \sigma, \tau, w, \psi)$ in the base manifold $\mathcal{M}$ are standard tensor fields defined on the base space. Therefore, the horizontal lift operator ($\uparrow$) can be applied to each of them. This procedure is the means by which their components are calculated in terms of the total space connection $\widehat{\nabla}$, as will be detailed next.

The general expressions for these lifted fields are summarized here,
\begin{equation}
  \label{eq:Levantamiento}
  \begin{aligned}
    \widetilde{\beta}(\widehat{X}) &= \widehat{P}(\widehat{\nabla}_{\widetilde{X}} \widehat{\eta}) ,
    & 
      \widetilde{\beta}^\nu{}_\mu &= \widehat{P}^\lambda{}_\mu \widehat{P}^\nu{}_\alpha \widehat{\nabla}_\lambda \widehat{\eta}^\alpha ,
    \\[1ex] 
    \widetilde{\gamma}(\widehat{X}) &= \widehat{P}(\widehat{\nabla}_{\widehat{\eta}} \widetilde{X} - \pounds_{\widehat{\eta}} \widetilde{X}) ,
    & 
      \widetilde{\gamma}^\nu{}_\mu &= \widetilde{\beta}^\nu{}_\mu + 2 \, \widehat{P}^\nu{}_\beta \, \widehat{P}^\alpha{}_\mu \, \widehat{\eta}^\lambda \, \widehat{\Gamma}_{[\lambda}{}^{\beta}{}_{\alpha]} ,
    \\[1ex]
    \widetilde{h}( \widehat{X}, \widehat{Y}) &= \widehat{\theta}(\widehat{\nabla}_{\widetilde{X}} \widetilde{Y}) ,
    & 
      h_{\mu\nu} &= - \widehat{P}^\beta{}_\mu\widehat{\theta}_{\lambda}\widehat{\nabla}_{\beta}(\widehat{\theta}_{\nu}\widehat{\eta}^{\lambda}) ,
    \\[1ex]
    \widetilde{\tau}(\widehat{X}) &= \widehat{\theta}(\widehat{\nabla}_{\widetilde{X}} \widehat{\eta}) ,
    & 
      \widetilde{\tau}_\mu &= \widehat{P}^\lambda{}_\mu \widehat{\theta}_\nu \widehat{\nabla}_\lambda \widehat{\eta}^\nu ,
    \\[1ex]
    \widetilde{w}(\widehat{X}) &= \widehat{\theta}(\widehat{\nabla}_{\widehat{\eta}} \widetilde{X}) ,
    & 
      \widetilde{w}_\nu &= -\widehat{\eta}^\beta \widehat{\theta}_\lambda \widehat{\nabla}_\beta(\widehat{\theta}_\nu \widehat{\eta}^\lambda) ,
    \\[1ex]
    \widetilde\psi &= \widehat{\theta}(\widehat{\nabla}_{\widehat{\eta}} \widehat{\eta}) ,
    & 
      \widetilde\psi &= \widehat{\eta}^\lambda \widehat{\theta}_\alpha \widehat{\nabla}_\lambda \widehat{\eta}^\alpha ,
    \\[1ex]
    \widetilde\sigma &= \widehat{P}(\widehat{\nabla}_{\widehat{\eta}} \widehat{\eta}) ,
    & 
      \widetilde\sigma^\mu &= \widehat{P}^\mu{}_\beta \widehat{\eta}^\lambda \widehat{\nabla}_\lambda \widehat{\eta}^\beta.
  \end{aligned}  
\end{equation}

In the particular case of vector fields, and with respect to an arbitrary basis $\widehat{e}_\mu$, this projector $\widehat{P}^\mu{}_\nu = \widehat{e}^\mu \widehat{P} \widehat{e}_\nu$ takes the explicit form:
\begin{equation}
  \widehat{P}^\mu{}_\nu = \delta^\mu{}_\nu - \widehat{\eta}^\mu \widehat{\theta}_\nu.
  \label{eq:ProjectorComponentsExplicit}
\end{equation}

These general formulas can be evaluated on a specific basis to obtain more concrete expressions. From now on, we will work directly in the basis that satisfies condition~\eqref{D-1Condicion}, that is, where $\widehat{\eta}^\nu = \delta^\nu_D$ and $\widehat{\theta}_D = 1$.

To illustrate the procedure, let us consider the field $\widetilde{\beta}$. The derivation of its components, $\widetilde{\beta}^k{}_i$, starts from the general expression in the table. The calculation is simplified by noting that, under the basis conditions, the covariant derivative reduces to $\widehat{\nabla}_\lambda \widehat{\eta}^\alpha = \widehat{\Gamma}_{\lambda}{}^{\alpha}{}_D$ and the projector $\widehat{P}^k{}_\alpha$ becomes $\delta^k{}_\alpha$. The explicit derivation is as follows:
\begin{equation}
  \begin{aligned}
    \widetilde{\beta}^k{}_i
    &= \left( \widehat{P}^\lambda{}_i \widehat{P}^k{}_\alpha \right) \widehat{\Gamma}_{\lambda}{}^{\alpha}{}_{D} \\
    &= (\widehat{P}^\lambda{}_i) \widehat{\Gamma}_{\lambda}{}^{k}{}_{D} \\
    &= \widehat{P}^j{}_i \widehat{\Gamma}_{j}{}^{k}{}_{D} + \widehat{P}^D{}_i \widehat{\Gamma}_{D}{}^{k}{}_{D} \\
    &= \delta^j{}_i \widehat{\Gamma}_{j}{}^{k}{}_{D} - \widehat{\theta}_i \widehat{\Gamma}_{D}{}^{k}{}_{D} \\
    &= \widehat{\Gamma}_{i}{}^{k}{}_{D} - \widehat{\theta}_i \widehat{\Gamma}_{D}{}^{k}{}_{D}. 
  \end{aligned}
  \label{eq:beta_component_form_final}
\end{equation}

Applying this same procedure to the other fields yields the complete set of their components,
\begin{equation}
  \begin{aligned}
    \widetilde{\beta}^k{}_i &= \widehat{\Gamma}_{i}{}^{k}{}_{D} - \widehat{\theta}_i \widehat{\Gamma}_{D}{}^{k}{}_{D}, \\
    \widetilde{\gamma}^k{}_j &= \widehat{\Gamma}_{D}{}^{k}{}_{j} - \widehat{\theta}_j \widehat{\Gamma}_{D}{}^{k}{}_{D}, \\
    \widetilde{h}_{ij} &= \widehat{\Gamma}_{i}{}^{D}{}_{j} - \partial_i \widehat{\theta}_j + \widehat{\theta}_k \widehat{\Gamma}_{i}{}^{k}{}_{j} - \widehat{\theta}_i \widehat{\Gamma}_{D}{}^{D}{}_{j} - \widehat{\theta}_j \widehat{\Gamma}_{i}{}^{D}{}_{D} - \widehat{\theta}_i \widehat{\theta}_j \widehat{\Gamma}_{D}{}^{D}{}_{D} \\
                            &\quad - \widehat{\theta}_i \widehat{\theta}_k \widehat{\Gamma}_{D}{}^{k}{}_{j} + \widehat{\theta}_j \widehat{\theta}_k \widehat{\Gamma}_{i}{}^{k}{}_{D} + \widehat{\theta}_i \widehat{\theta}_j \widehat{\theta}_k \widehat{\Gamma}_{D}{}^{k}{}_{D}, \\
    \widetilde{\psi} &= \widehat{\theta}_k \widehat{\Gamma}_{D}{}^{k}{}_{D} + \widehat{\Gamma}_{D}{}^{D}{}_{D}, \\
    \widetilde{\tau}_i &= \widehat{\theta}_k \widehat{\Gamma}_{i}{}^{k}{}_{D} + \widehat{\Gamma}_{i}{}^{D}{}_{D} - \widehat{\theta}_i (\widehat{\theta}_k \widehat{\Gamma}_{D}{}^{k}{}_{D} + \widehat{\Gamma}_{D}{}^{D}{}_{D}), \\
    \widetilde{w}_j &= \widehat{\theta}_k \widehat{\Gamma}_{D}{}^{k}{}_{j} + \widehat{\Gamma}_{D}{}^{D}{}_{j} - \widehat{\theta}_j (\widehat{\theta}_k \widehat{\Gamma}_{D}{}^{k}{}_{D} + \widehat{\Gamma}_{D}{}^{D}{}_{D}), \\
    \widetilde{\sigma}^k &= \widehat{\Gamma}_{D}{}^{k}{}_{D}.
  \end{aligned}
  \label{eq:camposLevantadosExpresion}
\end{equation}

We can now express the components of the affine connection in terms of the geometric objects defined on the base $\mathcal{M}$, in terms of their corresponding horizontal liftings to the total space $\widehat{\mathcal{M}}$, using an arbitrary basis $\widehat{e}_\mu$,
\begin{equation}
  \label{AnzatConeccion}
  \begin{aligned}
    \widehat{\Gamma}_{i}{}^{k}{}_{j} &= \Gamma_{i}{}^{k}{}_{j} + \widehat{\theta}_{j} \beta^{k}{}_{i} + \widehat{\theta}_{i} \gamma^{k}{}_{j} + \widehat{\theta}_{i} \widehat{\theta}_{j} \sigma^{k}, \\
    \widehat{\Gamma}_{D}{}^{k}{}_{j} &= \gamma^{k}{}_{j} + \widehat{\theta}_{j} \sigma^{k}, \\
    \widehat{\Gamma}_{i}{}^{D}{}_{j} &= h_{i j} + \widehat{\theta}_{i} w_{j} + \widehat{\theta}_{j} \tau_{i} + \partial_{i} \widehat{\theta}_{j} + \widehat{\theta}_{i} \widehat{\theta}_{j} \psi - \widehat{\theta}_{k} \widehat{\Gamma}_{i}{}^{k}{}_{j}, \\
    \widehat{\Gamma}_{i}{}^{k}{}_{D} &= \beta^{k}{}_{i} + \widehat{\theta}_{i} \sigma^{k}, \\
    \widehat{\Gamma}_{D}{}^{k}{}_{D} &= \sigma^{k}, \\
    \widehat{\Gamma}_{i}{}^{D}{}_{D} &= \tau_{i} + \widehat{\theta}_{i} \psi - \widehat{\theta}_{k} \beta^{k}{}_{i} - \widehat{\theta}_{i} \widehat{\theta}_{k} \sigma^{k}, \\
    \widehat{\Gamma}_{D}{}^{D}{}_{j} &= w_{j} + \widehat{\theta}_{j} \psi - \widehat{\theta}_{k} \gamma^{k}{}_{j} - \widehat{\theta}_{j} \widehat{\theta}_{k} \sigma^{k}, \\
    \widehat{\Gamma}_{D}{}^{D}{}_{D} &= \psi - \widehat{\theta}_{i} \sigma^{i}.
  \end{aligned}
\end{equation}
The results obtained show how these tensors are explicitly related to the derivatives and components of the connection $\widehat{\nabla}$, respecting the structure of the principal bundle.

It is very instructive to see how some of the key components of the ansatz arise directly from the geometry of the formalism. A conceptual derivation method, based on the geometric relation $\hat{e}^D = \widehat{\theta} - \widetilde{A}$ of Eq.~\eqref{A_Def}, illustrates the structure of the final result. For example, the component $\widehat{\Gamma}_{D}{}^{D}{}_{D}$ can be derived directly from its definition, as follows
\begin{align*}
  \widehat{\Gamma}_{D}{}^{D}{}_{D}
  &= \hat{e}^D ( \widehat{\nabla}_{\hat{e}_D} \hat{e}_D ) \\
  &= ( \widehat{\theta} - \widetilde{A} ) ( \widehat{\nabla}_{\hat{e}_D} \hat{e}_D ) \\
  &= \widehat{\theta}(\widehat{\nabla}_{\hat{e}_D} \hat{e}_D) - \widetilde{A}(\widehat{\nabla}_{\hat{e}_D} \hat{e}_D) \\
  &= \psi - \widehat{\theta}_k \sigma^k.
\end{align*}
This calculation confirms the result, but also reveals how the structure of the ansatz is intrinsically linked to the definition of the fields $\psi$ and $\sigma$.

The Eq.~\eqref{AnzatConeccion} is the central result of this section, representing a one-to-one correspondence between the affine connections of the total and base space, and it will be the fundamental tool that will allow us to calculate the curvature and derive the field equations in the following sections.

\section{Curvature in the Affine Kaluza--Klein Theory \label{sec:ReduccionCurvatura}}

In this section, our goal is to use this ansatz to explicitly compute some curvature tensors. This will allow us to analyze how the curvature of the total space $\widehat{\mathcal{M}}$ relates to the components of the connection in the base space $\mathcal{M}$ and the fiber directions, providing deeper insight into the geometry of the dimensional reduction in affine theories.

The curvature tensor, \(\widehat{\mathrm{R}}_{\nu \mu}{}^{\lambda}{}_{\kappa}\), is a fundamental measure of curvature on a manifold. Its components are defined in terms of the affine connection $\widehat{\Gamma}$ as follows,
\begin{equation}
  \widehat{\mathrm{R}}_{\nu \mu}{}^{\lambda}{}_{\kappa} = \widehat{\partial}_{\nu} \widehat{\Gamma}_{\mu}{}^{\lambda}{}_{\kappa} - \widehat{\partial}_{\mu} \widehat{\Gamma}_{\nu}{}^{\lambda}{}_{\kappa} + \widehat{\Gamma}_{\nu}{}^{\lambda}{}_{\eta} \widehat{\Gamma}_{\mu}{}^{\eta}{}_{\kappa} - \widehat{\Gamma}_{\mu}{}^{\lambda}{}_{\eta} \widehat{\Gamma}_{\nu}{}^{\eta}{}_{\kappa},
\end{equation}
where $\widehat{\partial}_{\nu} = \frac{\partial}{\partial \widehat{x}^{\nu}}$. This tensor is the main character in the building of gravitational models and would be essential for the remaining of our analysis.

The Ricci tensor, denoted by $\widehat{\mathrm{R}}_{\mu\kappa}$, is obtained by contracting the Riemann curvature tensor. Specifically, it is defined as
\begin{equation}
  \widehat{\mathrm{R}}_{\mu\kappa} = \widehat{\mathrm{R}}_{\nu \mu}{}^{\nu}{}_{\kappa}.
\end{equation}
In General Relativity, the field equations can be expressed in such a way that the geometric properties of the spacetime are described by the Ricci tensor, while the distribution of matter/energy is given by a combination of the energy-momentum tensor and its trace.

In the original Kaluza--Klein model, the relevant field equations were the vanishing Ricci tensor of a five-dimensional spacetime. Such purely gravitational model yields and effective theory in four dimensions consisting of General Relativity coupled to an electromagnetic potential and a scalar field. Interestingly, the vanishing Ricci tensor is also relevant in other gravitational theories,\cite{overduin97_kaluz_klein} including the affine model of Einstein and Eddington,\cite{einstein23_theor_affin_field,einstein23_zur_affin_feldt,eddington24_mathem_theor_relat} and even supergravity.\cite{duff86_kaluz_klein_super}

The general decomposition of the curvature tensor is carried out following these steps.

\paragraph*{Index Decomposition:} The Riemann tensor indices $\widehat{\mathrm{R}}_{\nu \mu}{}^{\lambda}{}_{\kappa}$ were systematically separated to distinguish between components in the base space $\mathcal{M}$ (latin indices) and the fiber direction (index $D$), e.g. $\widehat{A}_{\nu} \widehat{B}^{\mu} = \widehat{A}_i B^j + \widehat{A}_D \widehat{B}^D$.

\paragraph*{Substitution of the Ansatz:} The ansatz for the affine connection, derived in Equation \eqref{AnzatConeccion}, was substituted into the expression for the Ricci tensor. The Ricci tensor was then fully expressed in terms of the induced connections ($\Gamma_{i}{}^{k}{}_{j}$) and the tensors $\beta$, $\gamma$, $h$, $\tau$, $w$, $\sigma$ and $\psi$ defined in Sec.~\ref{sec:Descomposicion}, along with their derivatives. As discussed in Section~\ref{sec:KK-metrico}, $\widehat{\partial} \widetilde{f} = \partial f$ when $f$ does not depend on the coordinate of the extra dimension. 

\paragraph*{Grouping of Terms:} The resulting terms were regrouped to express the Ricci tensor in reduced dimension as
\begin{equation}
  R_{n m}{}^{l}{}_{k} = \partial_{n} \Gamma_{m}{}^{l}{}_{k} - \partial_{m} \Gamma_{n}{}^{l}{}_{k} + \Gamma_{n}{}^{l}{}_{e} \Gamma_{m}{}^{e}{}_{k} - \Gamma_{m}{}^{l}{}_{e} \Gamma_{n}{}^{e}{}_{k} .
  \label{eq:RiemanIntriReduc}
\end{equation}
Note that we have used the symbol $R_{ij}$, instead of $\mathrm{R}_{ij}$ (which is defined as $\mathrm{R}_{ij} \equiv \downarrow\, \hat{\mathrm{R}}_{ij}$), because in general $R_{ij} \neq \mathrm{R}_{ij}$. This is due to the fact that projecting the curvature of the total space (denoted $\hat{\mathrm{R}}$) using the ``hat-removal'' operator defined in \eqref{eq:QuitarHat} is not equivalent to computing the intrinsic curvature by \eqref{eq:RiemanIntriReduc}. The covariant derivative associated with the connection for a vector, consistent with the definition in Sec.~\ref{sec:Descomposicion}, is
\begin{equation}
  \nabla_i V^j = \partial_i V^j + \Gamma_i{}^j{}_k V^k.
\end{equation}

Hence, the algorithm yields the decomposition of the Ricci tensor of the connection on the total space $\widehat{\mathcal{M}}$ in terms of objects defined on the base space $\mathcal{M}$, 
\begin{equation}
  \begin{aligned}
    \widehat{\mathrm{R}}_{i j }
    & = R_{i j}-\nabla_{i}{w_{j}}
      + A_{i} \nabla_{k}{\gamma^{k}\,_{j}}
      + A_{j} \nabla_{k}{\beta^{k}\,_{i}}
      - A_{j} \nabla_{i}{\beta^{k}\,_{k}}
      - A_{j} \nabla_{i}{\psi} - \beta^{k}\,_{i} h_{k j}
      + \beta^{k}\,_{k} h_{i j}
    \\
    & \quad + \gamma^{k}\,_{j} \nabla_{k}{A_{i}}
      - \gamma^{k}\,_{j} \nabla_{i}{A_{k}}
      - \gamma^{k}\,_{j} h_{i k}
      + \psi h_{i j}
      - \tau_{i} w_{j}
      + A_{i} A_{j} \nabla_{k}{\sigma^{k}}
      + A_{i} \beta^{k}\,_{k} w_{j}
    \\
    & \quad - A_{i} \sigma^{k} h_{k j}
      + 2 A_{j} \Gamma_{[k}\,^{l}\,_{i]} \beta^{k}\,_{l}
      - A_{j} \beta^{k}\,_{i} \tau_{k}
      + A_{j} \beta^{k}\,_{i} w_{k}
      + A_{j} \beta^{k}\,_{k} \tau_{i}
      + A_{j} \nabla_{k}{A_{i}} \sigma^{k}
    \\
    & \quad - A_{j} \nabla_{i}{A_{k}} \sigma^{k}
      - A_{j} \sigma^{k} h_{i k}
      + 2 A_{k} \Gamma_{[l}\,^{k}\,_{i]} \gamma^{l}\,_{j}
      + A_{i} A_{j} \beta^{k}\,_{k} \psi
      - A_{i} A_{j} \beta^{k}\,_{l} \gamma^{l}\,_{k}
    \\
    & \quad - A_{i} A_{j} \sigma^{k} \tau_{k}
      - 2A_{j} A_{k} \Gamma_{[i}\,^{k}\,_{l]} \sigma^{l},
  \end{aligned}
\end{equation}
\begin{equation}
  \begin{aligned}
    \widehat{\mathrm{R}}_{DD}
    & = \nabla_{i}{\sigma^{i}}
      + A_{i} \beta^{j}\,_{j} \sigma^{i}
      + A_{i} \gamma^{i}\,_{j} \sigma^{j}
      + A_{i} A_{j} \sigma^{i} \sigma^{j}
      + \beta^{i}\,_{i} \psi-\beta^{i}\,_{i} A_{j} \sigma^{j}
      + A_{i} \sigma^{i} \psi
    \\
    & \quad
      - 2A_{i} \sigma^{i} A_{j} \sigma^{j}
      - \gamma^{i}\,_{j} \beta^{j}\,_{i}
      - \gamma^{i}\,_{j} A_{i} \sigma^{j}
      - A_{i} \sigma^{j} \beta^{i}\,_{j}
      - \sigma^{i} \tau_{i}
      - \sigma^{i} A_{i} \psi
    \\
    & \quad
      + \sigma^{i} A_{k} \beta^{k}\,_{i}
      + \sigma^{i} A_{i} A_{k} \sigma^{k},
  \end{aligned}
\end{equation}
\begin{equation}
  \begin{aligned}
    \widehat{\mathrm{R}}_{i D}
    &= \nabla_{j}{\beta^{j}\,_{i}}
      -\nabla_{i}{\beta^{j}\,_{j}}
      -\nabla_{i}{\psi}+A_{i} \nabla_{j}{\sigma^{j}}
      -2\Gamma_{[i}\,^{j}\,_{k]} \beta^{k}\,_{j}
      -\beta^{j}\,_{i} \tau_{j}
      +\beta^{j}\,_{i} w_{j }
      +\beta^{j}\,_{j} \tau_{i}
    \\
    & \quad
      +\nabla_{j}{A_{i}} \sigma^{j}
      -\nabla_{i}{A_{j}} \sigma^{j}
      -\sigma^{j} h_{i j}
      +A_{i} \beta^{j}\,_{j} \psi
      -A_{i} \beta^{j}\,_{k} \gamma^{k}\,_{j}
      -A_{i} \sigma^{j} \tau_{j}
      -2A_{j} \Gamma_{[i}\,^{j}\,_{k]} \sigma^{k},
  \end{aligned}
\end{equation}
and
\begin{equation}
  \widehat{\mathrm{R}}_{D j} =
  \nabla_{i}{\gamma^{i}\,_{j}}
  +A_{j} \nabla_{i}{\sigma^{i}}
  +\beta^{i}\,_{i} w_{j}
  -\sigma^{i} h_{i j}
  +A_{j} \beta^{i}\,_{i} \psi
  -A_{j} \beta^{i}\,_{k} \gamma^{k}\,_{i}
  -A_{j} \sigma^{i} \tau_{i}.
\end{equation}
This decomposition remains stable under the following scaling transformations:
\begin{align}
  \widehat{\eta}   & \longrightarrow \phi_1\, \widehat{\eta}, \\
  \widehat{\theta} & \longrightarrow \phi_2\, \widehat{\theta}.
\end{align}

In fact, scaling \(\widehat{\eta}\) does not change the generated vertical subspace, and scaling \(\widehat{\theta}\) does not alter its kernel, that is, the horizontal subspace. However, the quantity \(\widehat{\theta}(\widehat{\eta})\) transforms as \(\phi_1 \phi_2\, \widehat{\theta}(\widehat{\eta})\). Therefore, by imposing the normalization condition
\begin{equation*}
  \widehat{\theta}(\widehat{\eta}) = 1,
\end{equation*}
we fix the combination \(\phi_1 \phi_2 = 1\), thus eliminating one degree of freedom. One scaling degree of freedom remains, which we can use to impose a second normalization condition.

To motivate this second condition, we recall that in the geometry of Riemannian submanifolds there are two equivalent ways to define the second fundamental form:
\begin{enumerate}
\item Measures how much the covariant derivative of a vector projects onto the normal direction.
\item Measures how the normal direction (or its orthogonal complement) varies when moving tangentially.
\end{enumerate}
In our case, the tensor \(h\), defined by \(h_{ij} = \widehat{\theta}(\widehat{\nabla}_{\widetilde{e}_i} \widetilde{e}_j)\), measures how much the covariant derivative projects onto the fiber direction, even if this direction is not necessarily normal. On the other hand, the quantity \(- P^\beta{}_j\, \widehat{\nabla}_i \widehat{\theta}_\beta\) measures how the direction of the horizontal space varies when moving tangentially.\cite{spivak99_compr_introd_differ_geomet_vol}

Thus, by analogy between the horizontal space and the orthogonal complement, we can require the condition
\begin{equation}
  h_{ij} = -\widehat{P}^\beta{}_i\, \widehat{\theta}_\lambda\, \widehat{\nabla}_\beta (\widehat{\theta}_j\, \widehat{\eta}^\lambda) = -\widehat{P}^\beta{}_i\, \widehat{\nabla}_\beta \widehat{\theta}_j.
\end{equation}
This definition is analogous to that suggested by No\v{z}i\v{c}ka.\cite{nozicka50_le_vecteur_affin_et} It follows that
\begin{equation*}
  \widehat{\theta}_j\,\widehat{P}^\beta{}_i\, \widehat{\theta}_\lambda\, \widehat{\nabla}_\beta \, \widehat{\eta}^\lambda = 0,
\end{equation*}
and since we want this to hold for every \(\widehat{\theta}_j\), we must impose the following
\begin{equation}
  \widetilde{\tau}_i = \widehat{P}^\beta{}_i\, \widehat{\theta}_\lambda\, \widehat{\nabla}_\beta \, \widehat{\eta}^\lambda = 0,
\end{equation}
which defines the second normalization condition, analogous to that proposed by Schouten.\cite{schouten54_ricci_calcul}

Now, the scaling is as follows:
\begin{align}
  \widehat{\eta} &\rightarrow \phi\, \widehat{\eta}, \\
  \widehat{\theta} &\rightarrow \phi^{-1}\, \widehat{\theta}.
\end{align}
Therefore,
\begin{equation}
  \widehat{P}^\beta{}_i\, \widehat{\theta}_\lambda\, \widehat{\nabla}_\beta (\phi\, \widehat{\eta}^\lambda) = 0.
\end{equation}

Expanding the covariant derivative and applying the projector, we get
\begin{equation}
  \begin{aligned}
    0 &= \widehat{P}^\beta{}_i\, \widehat{\nabla}_\beta \phi + \phi\, \widehat{P}^\beta{}_i\, \widehat{\theta}_\lambda \left( \partial_\beta \widehat{\eta}^\lambda + \widehat{\Gamma}_\beta{}^\lambda{}_\sigma\, \widehat{\eta}^\sigma \right) \\
      &= \widehat{P}^\beta{}_i\, \partial_\beta \phi + \phi\, \widehat{P}^\beta{}_i\, \widehat{\theta}_\lambda \left( \partial_\beta \widehat{\eta}^\lambda + \widehat{\Gamma}_\beta{}^\lambda{}_\sigma\, \widehat{\eta}^\sigma \right).
  \end{aligned}
\end{equation}
Using that \(\widehat{\eta}^\lambda = \delta^\lambda{}_D\), \(\widehat{\theta}_\lambda = \delta^D{}_\lambda\), and the projector 
\(\widehat{P}^\beta{}_i = \delta^\beta{}_i - \delta^\beta{}_D\, \widehat{\theta}_i\), we obtain
\begin{equation}
  \begin{aligned}
    0 &= \left( \delta^\beta{}_i - \delta^\beta{}_D\, \widehat{\theta}_i \right)\, \partial_\beta \phi 
        + \phi \left( \delta^\beta{}_i - \delta^\beta{}_D\, \widehat{\theta}_i \right) \left( \partial_\beta \widehat{\eta}^D + \widehat{\Gamma}_\beta{}^D{}_D \right) \\
      &= \partial_i \phi - \widehat{\theta}_i\, \partial_D \phi 
        + \phi \left(   \widehat{\Gamma}_i{}^D{}_D - \widehat{\theta}_i\, \widehat{\Gamma}_D{}^D{}_D \right).
  \end{aligned}
\end{equation}
Using the relation \(\partial_D \phi = 0\), the equation reduces to
\begin{equation}
  \partial_i \log \phi = \widehat{\Gamma}_i{}^D{}_D - \widehat{\theta}_i\, \widehat{\Gamma}_D{}^D{}_D,
\end{equation}
and since \(\partial_D \widehat{\theta}_j = 0\), the system has a solution. In conclusion, we require only one normalization condition to impose \(\tau = 0\).

To further simplify the expressions obtained, we adopt the following considerations
\begin{itemize}
\item \textit{Torsion-free connection}. We consider a connection free of torsion,\cite{blagojevic02_gravit_gauge_symmet} namely
  \begin{equation*}
    \widehat{\Gamma}_\mu{}^\nu{}_\lambda = \widehat{\Gamma}_\lambda{}^\nu{}_\mu.
  \end{equation*}
  According to Eq.~\eqref{eq:Levantamiento}, this implies:
  \begin{align}
    \beta^i{}_j &= \gamma^i{}_j, \\
    \tau_i &= w_i. \label{eq:tau-gamma}
  \end{align}
  
\item \textit{Autoparallel fiber}. Since \(\widehat{\eta}\) is tangent to the fiber, the application of the horizontal projector \(P\) cancels its horizontal component:
  \begin{equation*}
    \widehat{\nabla}_{\widehat{\eta}} \widehat{\eta} = \alpha\, \widehat{\eta} \quad \Rightarrow \quad P(\widehat{\nabla}_{\widehat{\eta}} \widehat{\eta}) = 0 \quad \Rightarrow \quad \sigma = 0.
  \end{equation*}
  
\item \textit{Second normalization condition}. We impose that the covector \(\tau\) vanishes, namely \(\tau_i = 0\). From Eq.~\eqref{eq:tau-gamma}, it follows that \(w_i = 0\). 
  
\item \textit{Symmetric Ricci tensor}. In the reduced manifold, the Ricci tensor satisfies \(R_{ij} = R_{ji}\). This condition is equivalent to restricting to equiaffine connections.
\end{itemize}

With these considerations, the components of the Ricci tensor in the total space can be expressed as
\begin{equation}
  \begin{aligned}
    \widehat{\mathrm{R}}_{i j}
    & = R_{i j}+A_{i} \nabla_{k}{\beta^{k}\,_{j}}+A_{j} \nabla_{k}{\beta^{k}\,_{i}}-A_{j} \nabla_{i}{\beta^{k}\,_{k}}-A_{j} \nabla_{i}{\psi}- \beta^{k}\,_{i} h_{k j}+\beta^{k}\,_{j} \nabla_{k}{A_{i}}
    \\
    & \quad -\beta^{k}\,_{j} \nabla_{i}{A_{k}}-\beta^{k}\,_{j} h_{i k}+\beta^{k}\,_{k} h_{i j}+\psi h_{i j}+A_{i} A_{j} \beta^{k}\,_{k} \psi-A_{i} A_{j} \beta^{k}\,_{l} \beta^{l}\,_{k}
    \\
    & = 0,
  \end{aligned}
  \label{eq:Rij}
\end{equation}
\begin{equation}
  \widehat{\mathrm{R}}_{D j}
  = \nabla_{i}{\beta^{i}\,_{j}}+A_{j} \beta^{i}\,_{i} \psi-A_{j} \beta^{i}\,_{k} \beta^{k}\,_{i}
  = 0,
  \label{eq:RDi}
\end{equation}
\begin{equation}
  \widehat{\mathrm{R}}_{i D}
  = \nabla_{j}{\beta^{j}\,_{i}}-\nabla_{i}{\beta^{j}\,_{j}}-\nabla_{i}{\psi}+A_{i} \beta^{j}\,_{j} \psi-A_{i} \beta^{j}\,_{k} \beta^{k}\,_{j}
  = 0,
  \label{eq:RiD}
\end{equation}
and
\begin{equation}
  \widehat{\mathrm{R}}_{D D} = \beta^{i}\,_{i} \psi-\beta^{i}\,_{j} \beta^{j}\,_{i}
  = 0.
  \label{eq:RDD}
\end{equation}
Substituting \eqref{eq:RDD} into \eqref{eq:RDi}, we obtain
\begin{equation}
  \nabla_i \beta^i{}_j = 0,
  \label{eq:RdjSim}
\end{equation}
and replacing \eqref{eq:RdjSim} and \eqref{eq:RDi} in \eqref{eq:RiD}, we conclude that
\begin{equation}
  \beta^i{}_i + \psi = c_1,
\end{equation}
where \(c_1\) is a constant.

Next, we compute the antisymmetric part of \eqref{eq:Rij}. Using the previous results and substituting \eqref{eq:RDD} and \eqref{eq:RdjSim}, we find
\begin{equation}
  (\nabla_i A_j - \nabla_j A_i) c_1 = 0.
  \label{eq:RiiAntisim1}
\end{equation}

Recall that the electromagnetic field tensor, as discussed in Section~\ref{sec:KK-metrico}, is defined by
\begin{equation}
  F_{ij} = \partial_i A_j - \partial_j A_i = \nabla_i A_j - \nabla_j A_i.
\end{equation}
In order for Eq.~\eqref{eq:RiiAntisim1} to allow a nontrivial electromagnetic field, it is necessary that \(c_1 = 0\), which implies
\begin{equation}
  \beta^i{}_i = -\psi,
  \label{eq:RiDsim}
\end{equation}
and consequently, Eq.~\eqref{eq:RDD} can be rewritten as
\begin{equation}
  \beta^i{}_j \beta^j{}_i = -\psi^2.
  \label{eq:BB=psi}
\end{equation}

On the other hand, from Eq.~\eqref{eq:camposLevantadosExpresion}, we can separate the symmetric and antisymmetric parts of \(h_{ij}\) as follows
\begin{equation}
  h_{ij} = H_{ij} - \frac{1}{2} F_{ij},
\end{equation}
where \(H_{ij}\) is a symmetric tensor. By computing the symmetric part of Eq.~\eqref{eq:Rij}, and using the previous results, we obtain
\begin{equation}
  R_{ij} = \beta^k{}_{(i} F_{j)k} + 2 \beta^k{}_{(i} H_{j)k}.
  \label{eq:RijSimpli}
\end{equation}
Equations~\eqref{eq:RdjSim}, \eqref{eq:RiDsim}, \eqref{eq:BB=psi}, and \eqref{eq:RijSimpli} constitute the field equations of the reduced theory.

We observe that \(F\) does not exhibit intrinsic dynamics. However, to endow it with dynamics, we can take inspiration from the metric formulation of Kaluza--Klein, considering
\begin{equation}
  \widehat{\Gamma}_\mu{}^\lambda{}_\nu = \frac{1}{2} g^{\lambda\kappa} \left( \partial_\nu g_{\kappa\mu} + \partial_\mu g_{\kappa\nu} - \partial_\kappa g_{\mu\nu} \right).
\end{equation}
In this case, the expression for \(\beta\) according to \eqref{eq:camposLevantadosExpresion} is
\begin{equation}
  \beta^k{}_i = g^{kj} \frac{1}{2} \left( \partial_i A_j - \partial_j A_i \right) + \frac{1}{2} A^k \partial_i \phi + \frac{1}{2} A_i g^{kj} \partial_j \phi.
\end{equation}

For the case where \(\phi\) is constant (a situation that typically leads to inconsistencies), we obtain
\begin{equation}
  \beta^k{}_i = g^{kj} \frac{1}{2} F_{ij}.
  \label{Beta=F}
\end{equation}
Thus, according to equation \eqref{eq:RijSimpli}, it follows that
\begin{equation}
  R_{ij} = g^{kl} \frac{1}{2} F_{ik} F_{jl},
  \label{eq:RijRad}
\end{equation}
and equation \eqref{eq:RdjSim} becomes Maxwell’s equation
\begin{equation}
  \partial_k F^{kj} = 0.
\end{equation}
Condition \eqref{eq:RiDsim} requires \(\psi = 0\), which, when applied in Eq.~\eqref{eq:BB=psi}, leads to a condition on the electromagnetic Lorentz invariant,\cite{jackson99_class_elect}
\begin{equation}
  F^j{}_i F^i{}_j = 2\left( B^2 - E^2 \right) = 0,
\end{equation}
where \(E\) and \(B\) are the electric and magnetic field. This condition is satisfied whenever there are no sources (\(j_i = 0\)) and the Sommerfeld radiation conditions are met. That is, only radiative solutions are admitted.

Therefore, equation \eqref{eq:RijRad} effectively corresponds to the Einstein--Maxwell equations for purely radiative fields.

We have shown that, under the conditions of normalization and vanishing torsion, the dimensional reduction naturally leads to a consistent system of equations describing the gravitational and electromagnetic interaction in the base space.

\section{Conclusions \label{sec:Conclusiones}}

In this work, we have consider a bundle equipped with an Ehresmann connection and developed the appropriate decomposition for dimensional reduction in this context. The route guide of the formalism can be read in Table~\ref{Tab:desc}. The decomposition can be interpreted as a change of basis among adapted frames, according to Eq.~\eqref{eq:components-relations}. This perspective provides a clear geometric view of the relationship between the total and the reduced space.

Our method, inspired in the notion of parallel projections, is applicable to the traditional model of Kaluza--Klein, see Sec.~\ref{sec:KK-metrico}. The results obtained through our method generalize those of the metric decomposition a la Kaluza--Klein, and we found that the choice in Eq.~\eqref{D-1Condicion} makes the results coincide. Furthermore, such condition provides a natural geometric interpretation of the electromagnetic potential as the difference between the selected vector basis along the \emph{extra dimension}, as shown in Eq.~\eqref{A_Def}.

Applying Frobenius' theorem, we demonstrated that no submanifold of the total space can reproduce a nontrivial electromagnetic field. Consequently, the presence of an electromagnetic field necessarily implies that the horizontal distribution is non-integrable, confirming and generalizing the results reported by Li-Xin Li.\cite{li23_note_kaluz_klein_theor}

Subsequently, we carried out the decomposition of the affine connection in terms of objects defined on the reduced space. The use of adapted bases allowed us to establish a bijective correspondence between the connection in the total space and the objects in the reduced space, as reflected in equations \eqref{eq:camposLevantadosExpresion} and \eqref{AnzatConeccion}. We used these results to calculate some of the effective curvature tensor field on the reduced manifold.

The Ricci tensor plays a fundamental role in purely affine models of gravity, since vanishing Ricci is part of the space of solutions of these models, just as in the case of vacuum solutions to the field equations of General Relativity. For that reason, we computed explicitly the Ricci tensor projected onto the reduced space.

For the sake of concreteness, we focus on a simplified model, obtained after imposing certain requirements: (i) The affine connection of symmetric, i.e., without torsion; (ii) The vector along the direction of the fiber is self-parallel; (iii) The second normalization condition, which implies that both \(\tau_i\) and \(\sigma_i\) vanish; and, (iv) The affine connection on the reduced model is equiaffine, i.e., the Ricci tensor field is symmetric. The effective field equations on the reduced space are Eqs.~\eqref{eq:Rij}, \eqref{eq:RDi}, \eqref{eq:RiD} and~\eqref{eq:RDD}. These are the affine equivalent of the Kaluza--Klein equations.

We should notice that Eq.~\eqref{eq:Rij} is a simile of Einstein's equations, written in Ricci form. Explicitly,
\begin{equation*}
  \begin{aligned}
    R_{i j}
    & =
      - A_{i} \nabla_{k}{\beta^{k}\,_{j}}
      - A_{j} \nabla_{k}{\beta^{k}\,_{i}}
      + A_{j} \nabla_{i}{\beta^{k}\,_{k}}
      + A_{j} \nabla_{i}{\psi}
      + \beta^{k}\,_{i} h_{k j}
      - \beta^{k}\,_{j} \nabla_{k}{A_{i}}
    \\
    & \quad
      + \beta^{k}\,_{j} \nabla_{i}{A_{k}}
      + \beta^{k}\,_{j} h_{i k}
      - \beta^{k}\,_{k} h_{i j}
      - \psi h_{i j}
      - A_{i} A_{j} \beta^{k}\,_{k} \psi
      + A_{i} A_{j} \beta^{k}\,_{l} \beta^{l}\,_{k}.
  \end{aligned}
\end{equation*}
The right-hand side of the previous equation is an affine analogous of an energy-momentum-ish tensor in General Relativity.

Equations \eqref{eq:RDi} and \eqref{eq:RiD} resemble the Maxwell (or Yang--Mills) field equations for a field strength \(\beta^i{}_j\), while from Eq.~\eqref{eq:RiDsim} one could interpret that (stretching the analogy), the field \(\psi\) measures the failure of the field strength to be traceless, impacting the conformal properties of electrodynamics. However, in the particular scenario with \(\psi = 0\), the effective field equations correspond to those of an Einstein--Maxwell system.

One could argue that the identification in Eq.~\eqref{Beta=F}, \(\beta^i{}_j = F^i{}_j\), is responsible for introducing dynamics for the electromagnetic field. In electrodynamics, such a dynamics requires the existence of a metric (otherwise one cannot raise the index of the field strength tensor). Therefore, our identification supports the hypothesis that the metric might not be a fundamental object, but rather an emerging effect. In fact, Eq.~\eqref{Beta=F} matches the antisymmetric part of the Weingarten equation, which relates the second fundamental form to the extrinsic curvature in the submanifold geometry.

Although we initially did not assume the existence of a metric, we showed that a symmetric tensor \(H_{ij} = h_{(ij)}\) naturally arises, which can be interpreted as an emergent metric induced from the connection.

Looking ahead, we foresee two main lines of work: on the one hand, we intend to apply this dimensional reduction formalism to the study of affine theories, such as affine polynomial gravity, with the aim of incorporating matter fields, similarly to the success achieved in Kaluza--Klein theory. On the other hand, we observe that in the particular case where the horizontal distribution is integrable, the formalism becomes analogous to that of Arnowitt--Deser--Misner (also known as ADM formalism), opening the possibility of applying Hamiltonian methods for the study of conserved charges in affine theories, an avenue we plan to explore in future work.

\begin{acknowledgments}
  The work of O.C.-F. has been supported by ANID PIA/APOYO AFB230003 (Chile) and FONDECYT Grant 1230110 (Chile). J.V.-S. acknowledges the Doctoral scholarship of Universidad T\'ecnica Federico Santa Mar\'ia, granted by the Graduate Studies and the incentive program of the Pontificia Universidad Cat\'olica de Valpara\'iso.

  We would like to thank to the community of maintainers of the CADABRA software,\cite{peeters07_cadab,peeters07_introd_cadab,peeters07_symbol_field_theor_with_cadab,brewin09_rieman_normal_coord_expan,peeters18_cadab,brewin19_using_cadab_tensor_comput_gener_relat,kulyabov19_new_featur_secon_version_cadab,price22_hidin_canon_tensor_comput_algeb,castillo-felisola22_cadab_python_algor_gener_i,castillo-felisola22_cadab_python_algor_gener_ii} which we used to cross-check the algebraic manipulations of our formalism.
\end{acknowledgments}

\appendix

\section{Calculations for the Connection Decomposition \label{Calculations_Connection_Decomposition}}

This appendix details the explicit calculations for some of the key results presented in the main body of the article.

We begin by deriving the relationship between the components of the induced connection on the base manifold and the components of the affine connection on the total space.

Although the affine connection $\widehat{\nabla}$ itself is not a local tensorial object, therefore, does not possess a ``horizontal lift'' in the same sense as a tensor field, it is possible to establish a relationship between its components and those of the connection $\nabla$ induced on the base manifold $\mathcal{M}$. 
For this purpose, we use the adapted bases introduced in Section \ref{sec:Descomposicion}, specifically the basis $\mathcal{B} = \{ \widehat{e}_0, \widehat{e}_1, ..., \widehat{e}_D = \widehat{\eta} \}$ of the tangent space $T\widehat{\mathcal{M}}$ and its associated dual.

The components of the induced connection $\nabla$ in the base manifold $\mathcal{M}$, $\Gamma_i{}^k{}_j$, are defined from the projection of the covariant derivative of the horizontal base vectors:
\begin{align*}
  \Gamma_i{}^k{}_j &= e^k (\nabla_{e_i} e_j) \\
                   &= e^k ( \downarrow (\widehat{\nabla}_{\widetilde{e}_i} \widetilde{e}_j) ).
\end{align*}
As discussed (see, for example, the relation for covectors in Section \ref{sec:fibrados_conexion} and Eq. \eqref{eq:CovectorL} in the context of projections), the action of the base space covector $e^k$ on a vector projected from the total space can be expressed by lifting $e^k$ to the total space, $\widehat{e}^k$, acting on the horizontal lift of the argument. Thus, $e^k ( \downarrow (\widehat{\nabla}_{\widetilde{e}_i} \widetilde{e}_j) ) = \widehat{e}^k ( \uparrow ( \downarrow (\widehat{\nabla}_{\widetilde{e}_i} \widetilde{e}_j) ) )$.
Using the horizontal projector $\widehat{P} = \uparrow \circ \downarrow$, we obtain the following:
\begin{align*}
  \Gamma_i{}^k{}_j &= \widehat{e}^k \left( \widehat{P} (\widehat{\nabla}_{\widetilde{e}_i} \widetilde{e}_j) \right) \\
                   &= \widehat{e}^k \left( \widehat{\nabla}_{\widetilde{e}_i} \widetilde{e}_j - \widehat{\theta}(\widehat{\nabla}_{\widetilde{e}_i} \widetilde{e}_j) \widehat{\eta} \right).
\end{align*}
Since the base covectors $\widehat{e}^k$ are dual to the lifted horizontal base vectors $\widetilde{e}_k$ (and therefore orthogonal to $\widehat{\eta}$), we have $\widehat{e}^k(\widehat{\eta}) = 0$. The previous expression simplifies to:
\begin{align*}
  \Gamma_i{}^k{}_j &= \widehat{e}^k \left( \widehat{\nabla}_{\widetilde{e}_i} \widetilde{e}_j \right) \\
                   &= \widehat{e}^k \left( \widehat{\nabla}_{\widehat{e}_i - \widehat{\theta}_i \widehat{\eta}} (\widehat{e}_j - \widehat{\theta}_j \widehat{\eta}) \right) \\
                   &= \widehat{e}^k \Big( \widehat{\nabla}_{\widehat{e}_i} \widehat{e}_j - (\widehat{e}_i \widehat{\theta}_j) \widehat{\eta} - \widehat{\theta}_j \widehat{\nabla}_{\widehat{e}_i} \widehat{\eta} \\ 
                   & \qquad - (\widehat{\eta} \widehat{\theta}_i) \widehat{e}_j + \widehat{\theta}_i \widehat{\nabla}_{\widehat{\eta}} \widehat{e}_j + \widehat{\theta}_i (\widehat{\eta} \widehat{\theta}_j) \widehat{\eta} - \widehat{\theta}_i \widehat{\theta}_j \widehat{\nabla}_{\widehat{\eta}} \widehat{\eta} \Big).
\end{align*}
Considering that the components of the vector potential $A_i = \widehat{\theta}_i$ depend only on the base space coordinates $x^j$, $A_i = A_i(x^j)$), and the consistency with the \(G\)-invariance of the connection, $\partial_D \widehat{\Gamma}^\rho_{\mu\nu} = 0$ (see Eq. \eqref{eq:GInvarianzaConexion}), it follows that
$\widehat{\eta}(\widehat{\theta}_i) = \partial_D \widehat{\theta}_i = 0$.

This implies that the terms $(\widehat{\eta} \widehat{\theta}_i) \widehat{e}_j$ and $\widehat{\theta}_i (\widehat{\eta} \widehat{\theta}_j) \widehat{\eta}$ are null. and similarly, the term $(\widehat{e}_i \widehat{\theta}_j) \widehat{\eta}$ vanishes upon application of $\widehat{e}^k$, since $\widehat{e}^k(\widehat{\eta})=0$.

Therefore, applying $\widehat{e}^k$ and recalling the definition of the Christoffel coefficients $\widehat{\Gamma}_{\mu}{}^{\lambda}{}_{\nu}$, we obtain
\begin{equation}
  \begin{aligned}
    \Gamma_i{}^k{}_j
    &= \widehat{e}^k \left( \widehat{\nabla}_{\widehat{e}_i} \widehat{e}_j - \widehat{\theta}_j \widehat{\nabla}_{\widehat{e}_i} \widehat{\eta} - \widehat{\theta}_i \widehat{\nabla}_{\widehat{\eta}} \widehat{e}_j + \widehat{\theta}_i \widehat{\theta}_j \widehat{\nabla}_{\widehat{\eta}} \widehat{\eta} \right) \\
    &= \widehat{\Gamma}_{i}{}^k{}_j - \widehat{\theta}_j \widehat{\Gamma}_{i}{}^k{}_D - \widehat{\theta}_i \widehat{\Gamma}_{D}{}^k{}_j + \widehat{\theta}_i \widehat{\theta}_j \widehat{\Gamma}_{D}{}^k{}_D. \label{eq:GammaDs}
  \end{aligned}
\end{equation}

On the other hand, we could also start from the expression for the projected \(1\)-form $\Gamma^k{}_j$ given in Eq.~\eqref{eq:AltProjectedConnection}, and obtain the components $\Gamma_i{}^k{}_j$ through the projection to the base vectors $e_i$, as follows
\begin{align*}
  \Gamma_i{}^k{}_j
  &= \Gamma^k{}_j (e_i) \\
  &= \widehat{P}^k{}_\kappa \widehat{P}^\nu{}_j \widetilde{\Gamma}^\kappa{}_\nu (\widehat{e}_i) \\
  &= \widehat{P}^k{}_\kappa \widehat{P}^\nu{}_j \widehat{\Gamma}^\kappa{}_\nu (\widetilde{e}_i) \\
  &= \widehat{P}^k{}_\kappa \widehat{P}^\nu{}_j \widehat{\Gamma}_{\lambda}{}^\kappa{}_\nu \widehat{e}^\lambda (\widetilde{e}_i) \\
  &= \widehat{P}^k{}_\kappa \widehat{P}^\nu{}_j \left( \widehat{\Gamma}_{l}{}^\kappa{}_\nu \widehat{e}^l(\widetilde{e}_i) + \widehat{\Gamma}_{D}{}^\kappa{}_\nu \widehat{e}^D(\widetilde{e}_i) \right).
\end{align*}
Now, the condition~\eqref{D-1Condicion} implies that $\widehat{\eta}^\nu = \delta^\nu_D$ and $\widehat{\theta}_D = 1$. Then, 
\begin{equation}
\widehat{e}^D(\widetilde{e}_j) = \widehat{e}^D(\widehat{e}_j - \widehat{\theta}_j \widehat{\eta}) = -\widehat{\theta}_j \widehat{e}^D(\widehat{\eta}) = -\widehat{\theta}_j \widehat{e}^D(\widehat{e}_D) = -\widehat{\theta}_j.
\end{equation}
 Also, from Eq.~\eqref{eq:components-relations}, we have $\widetilde{e}_l = \widehat{e}_l$, thus
\begin{align*}
  \Gamma_i{}^k{}_j
  &= \widehat{P}^k{}_\kappa \widehat{P}^\nu{}_j \left( \widehat{\Gamma}_{i}{}^\kappa{}_\nu - \widehat{\theta}_i \widehat{\Gamma}_{D}{}^\kappa{}_\nu \right) \\
  &= (\delta^k{}_\kappa - \widehat{\eta}^k \widehat{\theta}_\kappa)(\delta^\nu{}_j - \widehat{\eta}^\nu \widehat{\theta}_j)(\widehat{\Gamma}_{i}{}^\kappa{}_\nu - \widehat{\theta}_i \widehat{\Gamma}_{D}{}^\kappa{}_\nu) \\
  &= \widehat{\Gamma}_{i}{}^k{}_j - \widehat{\theta}_j \widehat{\Gamma}_{i}{}^k{}_D - \widehat{\theta}_i \widehat{\Gamma}_{D}{}^k{}_j + \widehat{\theta}_i \widehat{\theta}_j \widehat{\Gamma}_{D}{}^k{}_D.
\end{align*}

We have revealed the equivalence of the procedures to obtain the coefficients of the induced connection on the manifold \(\mathcal{M}\).

Now, we present the detailed calculation of the components of the tensor $\widetilde{\gamma}$, defined by $\widetilde{\gamma}(\widehat{X}) = \widehat{P}\big(\widehat{\nabla}_{\hat{\eta}} \widetilde{X} - \pounds_{\hat{\eta}} \widetilde{X}\big)$. In a coordinate basis $\{\widehat{e}_\mu\}$, the component $\widetilde{\gamma}^{\mu}{}_{\nu}$ is obtained by projecting the result onto $\widehat{e}_\mu$:
\begin{align*}
  \tilde{\gamma}^{\mu}{}_{\nu}
  & = \hat{e}^{\mu} \left( \hat{P}\left( \hat{\nabla}_{\hat{\eta}} (\hat{P}^{\rho}{}_{\nu} \hat{e}_{\rho}) - \pounds_{\hat{\eta}} (\hat{P}^{\rho}{}_{\nu} \hat{e}_{\rho}) \right) \right).
\end{align*}
We apply the Leibniz rule for the covariant derivative and the Lie derivative to the terms involving $\hat{P}^{\rho}{}_{\nu}$. Recalling that $\pounds_{\hat{\eta}}(\hat{P}^{\rho}{}_{\nu}) = \hat{\eta}^{\lambda}\partial_{\lambda}\hat{P}^{\rho}{}_{\nu}$ (since $\hat{P}^{\rho}{}_{\nu}$ are scalar components and $\hat{\eta}$ is a vector field), we obtain:
\begin{equation*}
\tilde{\gamma}^{\mu}{}_{\nu} = \hat{e}^{\mu} \left( \hat{P}\big( (\hat{\eta}^{\lambda}\partial_{\lambda}\hat{P}^{\rho}{}_{\nu}) \hat{e}_{\rho} + \hat{P}^{\rho}{}_{\nu} \hat{\nabla}_{\hat{\eta}} \hat{e}_{\rho} - (\hat{\eta}^{\lambda}\partial_{\lambda}\hat{P}^{\rho}{}_{\nu}) \hat{e}_{\rho} - \hat{P}^{\rho}{}_{\nu} \pounds_{\hat{\eta}} \hat{e}_{\rho} \big) \right).
\end{equation*}
The first and third terms inside the parentheses cancel out. The expression simplifies to:
\begin{equation*}
\tilde{\gamma}^{\mu}{}_{\nu} = \hat{e}^{\mu} \left( \hat{P}\left( \hat{P}^{\rho}{}_{\nu} \left( \hat{\nabla}_{\hat{\eta}} \hat{e}_{\rho} - \pounds_{\hat{\eta}} \hat{e}_{\rho} \right) \right) \right).
\end{equation*}
We use the known relations: $\hat{\nabla}_{\hat{\eta}} \hat{e}_{\rho} = \hat{\eta}^{\lambda} \hat{\Gamma}_{\lambda}{}^{\sigma}{}_{\rho} \hat{e}_{\sigma}$ and $\pounds_{\hat{\eta}} \hat{e}_{\rho} = -\partial_{\rho} \hat{\eta}^{\sigma} \hat{e}_{\sigma}$. Substituting these into the previous expression:
\begin{align*}
\tilde{\gamma}^{\mu}{}_{\nu} &= \hat{e}^{\mu} \left( \hat{P}\left( \hat{P}^{\rho}{}_{\nu} \left( \hat{\eta}^{\lambda} \hat{\Gamma}_{\lambda}{}^{\sigma}{}_{\rho} \hat{e}_{\sigma} - (-\partial_{\rho} \hat{\eta}^{\sigma} \hat{e}_{\sigma}) \right) \right) \right) \\
&= \hat{e}^{\mu} \left( \hat{P}\left( \hat{P}^{\rho}{}_{\nu} \left( \hat{\eta}^{\lambda} \hat{\Gamma}_{\lambda}{}^{\sigma}{}_{\rho} + \partial_{\rho} \hat{\eta}^{\sigma} \right) \hat{e}_{\sigma} \right) \right).
\end{align*}
Applying the projector $\hat{P}$ (whose components are $\hat{P}^{\mu}{}_{\sigma}$) to the resulting vector, we obtain:
\begin{equation*}
\tilde{\gamma}^{\mu}{}_{\nu} = \hat{P}^{\mu}{}_{\sigma} \hat{P}^{\rho}{}_{\nu} \left( \hat{\eta}^{\lambda} \hat{\Gamma}_{\lambda}{}^{\sigma}{}_{\rho} + \partial_{\rho} \hat{\eta}^{\sigma} \right).
\end{equation*}
To express this result in terms of the covariant derivative of $\hat{\eta}^{\sigma}$, we use the identity $\partial_{\rho} \hat{\eta}^{\sigma} = \hat{\nabla}_{\rho} \hat{\eta}^{\sigma} - \hat{\eta}^{\lambda} \hat{\Gamma}_{\rho}{}^{\sigma}{}_{\lambda}$. Substituting
\begin{align*}
\tilde{\gamma}^{\mu}{}_{\nu} &= \hat{P}^{\mu}{}_{\sigma} \hat{P}^{\rho}{}_{\nu} \left( \hat{\eta}^{\lambda} \hat{\Gamma}_{\lambda}{}^{\sigma}{}_{\rho} + (\hat{\nabla}_{\rho} \hat{\eta}^{\sigma} - \hat{\eta}^{\lambda} \hat{\Gamma}_{\rho}{}^{\sigma}{}_{\lambda}) \right) \\
&= \hat{P}^{\mu}{}_{\sigma} \hat{P}^{\rho}{}_{\nu} \left( \hat{\eta}^{\lambda} (\hat{\Gamma}_{\lambda}{}^{\sigma}{}_{\rho} - \hat{\Gamma}_{\rho}{}^{\sigma}{}_{\lambda}) + \hat{\nabla}_{\rho} \hat{\eta}^{\sigma} \right).
\end{align*}
Finally, recognizing that $\hat{\Gamma}_{\lambda}{}^{\sigma}{}_{\rho} - \hat{\Gamma}_{\rho}{}^{\sigma}{}_{\lambda} = 2\hat{\Gamma}_{[\lambda}{}^{\sigma}{}_{\rho]} = \hat{T}_{\lambda\rho}{}^{\sigma}$ (where $\hat{T}$ is the torsion tensor), the expression for $\tilde{\gamma}^{\mu}{}_{\nu}$ is:
\begin{equation*}
\tilde{\gamma}^{\mu}{}_{\nu} = \hat{P}^{\mu}{}_{\sigma} \hat{P}^{\rho}{}_{\nu} \left( \hat{\eta}^{\lambda} \hat{T}_{\lambda\rho}{}^{\sigma} + \hat{\nabla}_{\rho} \hat{\eta}^{\sigma} \right).
\end{equation*}
Alternatively, using the antisymmetrization notation for the connection:
\begin{equation*}
\tilde{\gamma}^{\mu}{}_{\nu} = \hat{P}^{\mu}{}_{\sigma} \hat{P}^{\rho}{}_{\nu} \left( 2 \hat{\eta}^{\lambda} \hat{\Gamma}_{[\lambda}{}^{\sigma}{}_{\rho]} + \hat{\nabla}_{\rho} \hat{\eta}^{\sigma} \right).
\end{equation*}

Next, we calculate the auxiliary expression $\theta(\hat{\nabla}_{\hat{e}_{\beta}} (\widehat{P}^{\lambda}{}_{\nu} \hat{e}_{\lambda}))$. This appears recurrently, and its breakdown is instructive.
First, we apply the Leibniz rule for the covariant derivative on the product $\widehat{P}^{\lambda}{}_{\nu}\hat{e}_{\lambda}$:
\begin{align*}
\theta(\hat{\nabla}_{\hat{e}_{\beta}}(\widehat{P}^{\lambda}{}_{\nu}\hat{e}_{\lambda}))
&= \theta\left( (\partial_{\beta} \widehat{P}^{\lambda}{}_{\nu}) \widehat{e}_{\lambda} + \widehat{P}^{\lambda}{}_{\nu} \widehat{\nabla}_{\hat{e}_{\beta}} \widehat{e}_{\lambda} \right).
\end{align*}
We substitute the expression for the projector derivative, $\partial_{\beta} \widehat{P}^{\lambda}{}_{\nu} = -\partial_{\beta}(\hat{\theta}_{\nu} \widehat{\eta}^{\lambda})$, and the definition of the covariant derivative of a basis vector, $\widehat{\nabla}_{\hat{e}_{\beta}} \widehat{e}_{\lambda} = \widehat{\Gamma}_{\beta}{}^{\mu}{}_{\lambda} \widehat{e}_{\mu}$. Upon applying the 1-form $\hat{\theta}$ (which annihilates horizontal terms and returns $\hat{\theta}(\hat{e}_\mu) = \hat{\theta}_\mu$), we obtain:
\begin{align*}
\theta(\hat{\nabla}_{\hat{e}_{\beta}}(\widehat{P}^{\lambda}{}_{\nu}\hat{e}_{\lambda}))
&= \theta\left( -(\partial_{\beta}(\hat{\theta}_{\nu} \widehat{\eta}^{\lambda})) \widehat{e}_{\lambda} + \widehat{P}^{\lambda}{}_{\nu} \widehat{\Gamma}_{\beta}{}^{\mu}{}_{\lambda} \widehat{e}_{\mu} \right) \\
&= -(\partial_{\beta}(\hat{\theta}_{\nu} \widehat{\eta}^{\lambda})) \widehat{\theta}_{\lambda} + \widehat{P}^{\lambda}{}_{\nu} \widehat{\Gamma}_{\beta}{}^{\mu}{}_{\lambda} \widehat{\theta}_{\mu}.
\end{align*}
Now, we explicitly expand the projector $\widehat{P}^{\lambda}{}_{\nu} = \delta^{\lambda}{}_{\nu} - \widehat{\eta}^{\lambda} \widehat{\theta}_{\nu}$ in the second term:
\begin{align*}
\theta(\hat{\nabla}_{\hat{e}_{\beta}}(\widehat{P}^{\lambda}{}_{\nu}\hat{e}_{\lambda}))
&= -(\partial_{\beta}(\hat{\theta}_{\nu} \widehat{\eta}^{\lambda})) \widehat{\theta}_{\lambda} + (\delta^{\lambda}{}_{\nu} - \widehat{\eta}^{\lambda} \widehat{\theta}_{\nu}) \widehat{\Gamma}_{\beta}{}^{\mu}{}_{\lambda} \widehat{\theta}_{\mu} \\
&= -(\partial_{\beta}(\hat{\theta}_{\nu} \widehat{\eta}^{\lambda})) \widehat{\theta}_{\lambda} + \widehat{\Gamma}_{\beta}{}^{\mu}{}_{\nu} \widehat{\theta}_{\mu} - \widehat{\eta}^{\lambda} \widehat{\theta}_{\nu} \widehat{\Gamma}_{\beta}{}^{\mu}{}_{\lambda} \widehat{\theta}_{\mu}.
\end{align*}
To reveal the structure of a covariant derivative, it is useful to multiply the second term of this last expression by unity, conveniently written as $1=\hat{\eta}^\sigma \widehat{\theta}_\sigma$. This step, along with a relabeling of dummy indices and a rearrangement, allows the factorization of the common term $-\widehat{\theta}_{\lambda}$:
\begin{align*}
\theta(\hat{\nabla}_{\hat{e}_{\beta}}(\widehat{P}^{\lambda}{}_{\nu}\hat{e}_{\lambda}))
&= -(\partial_{\beta}(\hat{\theta}_{\nu} \widehat{\eta}^{\lambda})) \widehat{\theta}_{\lambda} + \widehat{\Gamma}_{\beta}{}^{\mu}{}_{\nu} \widehat{\theta}_{\mu}(\hat{\eta}^{\sigma}\hat{\theta}_{\sigma}) - \widehat{\eta}^{\lambda} \widehat{\theta}_{\nu} \widehat{\Gamma}_{\beta}{}^{\mu}{}_{\lambda} \widehat{\theta}_{\mu} \\[1ex]
&= - \widehat{\theta}_{\lambda} \left( \partial_{\beta}(\hat{\theta}_{\nu} \widehat{\eta}^{\lambda}) + \widehat{\Gamma}_{\beta}{}^{\lambda}{}_{\sigma}(\hat{\theta}_{\nu}\widehat{\eta}^{\sigma}) - \widehat{\Gamma}_{\beta}{}^{\sigma}{}_{\nu}(\hat{\theta}_{\sigma} \widehat{\eta}^{\lambda}) \right).
\end{align*}
This expression corresponds precisely to the covariant derivative of the object $\hat{\theta}_{\nu}\hat{\eta}^{\lambda}$:
\begin{equation}
\theta(\hat{\nabla}_{\hat{e}_{\beta}}(\widehat{P}^{\lambda}{}_{\nu}\hat{e}_{\lambda})) = - \widehat{\theta}_{\lambda} \widehat{\nabla}_{\beta}(\hat{\theta}_{\nu} \widehat{\eta}^{\lambda}). \label{eq:thetaExtra}
\end{equation}

With these results, we proceed to find the local components of the relevant geometric objects.

The tensor $\tilde{\beta}$, defined as $\tilde{\beta}(\tilde{X}) = \hat{P}(\hat{\nabla}_{\tilde{X}} \hat{\eta})$, has components $\tilde{\beta}^{\mu}{}_{\nu}$:
\begin{align*}
\tilde{\beta}^{\mu}{}_{\nu} &= \hat{e}^{\mu} \left( \hat{P}(\hat{\nabla}_{P\hat{e}_{\nu}} \hat{\eta}) \right) \\
&= \hat{e}^{\mu} \left( \hat{P}(\hat{\nabla}_{P^{\lambda}{}_{\nu}\hat{e}_{\lambda}} \hat{\eta}) \right) \\
&= P^{\lambda}{}_{\nu} \, \hat{e}^{\mu} \left( \hat{P}(\hat{\nabla}_{\hat{e}_{\lambda}} \hat{\eta}) \right) \\
&= P^{\lambda}{}_{\nu} \, \hat{e}^{\mu} \left( \hat{P}((\hat{\nabla}_{\lambda} \hat{\eta}^{\alpha})\hat{e}_{\alpha}) \right) \\
&= P^{\lambda}{}_{\nu} (\hat{\nabla}_{\lambda} \hat{\eta}^{\alpha}) \hat{P}^{\mu}{}_{\alpha}.
\end{align*}

The horizontal 1-form $\tilde{\tau}$, defined by $\tilde{\tau}(\tilde{X}) = \hat{\theta}(\hat{\nabla}_{\tilde{X}}\hat{\eta})$, has components $\tilde{\tau}_{\mu}$:
\begin{align*}
\tilde{\tau}_{\mu} &= \hat{\theta}(\hat{\nabla}_{P\hat{e}_{\mu}}\hat{\eta}) \\
&= \hat{\theta}(\hat{\nabla}_{P^{\lambda}{}_{\mu}\hat{e}_{\lambda}}\hat{\eta}) \\
&= P^{\lambda}{}_{\mu} \, \hat{\theta}( \hat{\nabla}_{\hat{e}_{\lambda}}\hat{\eta}) \\
&= P^{\lambda}{}_{\mu} \, \hat{\theta}( (\hat{\nabla}_{\lambda}\hat{\eta}^{\nu})\hat{e}_{\nu}) \\
&= P^{\lambda}{}_{\mu} (\hat{\nabla}_{\lambda}\hat{\eta}^{\nu}) \hat{\theta}_{\nu}.
\end{align*}

The 1-form $\tilde{w}$, defined as $\tilde{w}(\tilde{X}) = \hat{\theta}(\hat{\nabla}_{\hat{\eta}}\tilde{X})$, has components $\tilde{w}_{\nu}$ that are calculated using \eqref{eq:thetaExtra}:
\begin{align*}
\tilde{w}_{\nu} &= \theta(\hat{\nabla}_{\hat{\eta}}(P\hat{e}_{\nu})) \\
&= \hat{\eta}^{\beta} \theta(\hat{\nabla}_{\hat{e}_{\beta}}(P^{\lambda}{}_{\nu}\hat{e}_{\lambda})) \\
&= -\hat{\eta}^{\beta} \hat{\theta}_{\lambda} \hat{\nabla}_{\beta}(\hat{\theta}_{\nu}\hat{\eta}^{\lambda}).
\end{align*}

The tensor $\tilde{h}$, defined by $\tilde{h}(\tilde{X}, \tilde{Y}) = \hat{\theta}(\hat{\nabla}_{\tilde{X}} \tilde{Y})$, has components $h_{\mu\nu}$:
\begin{align*}
h_{\mu\nu} &= \hat{\theta}(\hat{\nabla}_{P\hat{e}_{\mu}}P\hat{e}_{\nu}) \\
&= \hat{\theta}(\hat{\nabla}_{P^{\beta}{}_{\mu}\hat{e}_{\beta}}(P^{\lambda}{}_{\nu}\hat{e}_{\lambda})) \\
&= P^{\beta}{}_{\mu} \, \hat{\theta}(\hat{\nabla}_{\hat{e}_{\beta}}(P^{\lambda}{}_{\nu}\hat{e}_{\lambda})) \\
&= -P^{\beta}{}_{\mu} \hat{\theta}_{\lambda} \hat{\nabla}_{\beta}(\hat{\theta}_{\nu}\hat{\eta}^{\lambda}).
\end{align*}

The scalar field $\tilde{\psi}$, which corresponds to the vertical projection of the self-acceleration of $\hat{\eta}$, $\tilde{\psi} = \hat{\theta}(\hat{\nabla}_{\hat{\eta}} \hat{\eta})$, is written as:
\begin{align*}
\tilde{\psi} &= \hat{\eta}^{\lambda} \hat{\theta} (\hat{\nabla}_{\hat{e}_{\lambda}} \hat{\eta}) \\
&= \hat{\eta}^{\lambda} \hat{\theta} ((\hat{\nabla}_{\lambda} \hat{\eta}^{\alpha}) \hat{e}_{\alpha}) \\
&= \hat{\eta}^{\lambda} (\hat{\nabla}_{\lambda} \hat{\eta}^{\alpha}) \hat{\theta}_{\alpha}.
\end{align*}

The vector field $\tilde{\sigma}$, which is the horizontal part of the self-acceleration of $\hat{\eta}$, $\tilde{\sigma} = \hat{P}(\hat{\nabla}_{\hat{\eta}}\hat{\eta})$, has components $\tilde{\sigma}^{\mu}$:
\begin{align*}
\tilde{\sigma}^{\mu} &= \hat{e}^{\mu} \left( \hat{P} (\hat{\nabla}_{\hat{\eta}}\hat{\eta}) \right) \\
&= \hat{\eta}^{\lambda} \hat{e}^{\mu} \left( \hat{P}( \hat{\nabla}_{\hat{e}_{\lambda}} \hat{\eta}) \right) \\
&= \hat{\eta}^{\lambda} \hat{e}^{\mu} \left( \hat{P}( (\hat{\nabla}_{\lambda}\hat{\eta}^{\beta})\hat{e}_{\beta}) \right) \\
&= \hat{\eta}^{\lambda} (\hat{\nabla}_{\lambda}\hat{\eta}^{\beta}) \hat{P}^{\mu}{}_{\beta}.
\end{align*}

In Eq.~\eqref{eq:Levantamiento}, it is shown how the objects originally defined on $\mathcal{M}$ can be completely reexpressed as a function of quantities of the total space $\widehat{\mathcal{M}}$, preserving the structure of the bundle and making explicit the action of the affine connection $\widehat{\nabla}$.

\bibliographystyle{apsrev4-2}
\bibliography{references}
\end{document}